%% file: CSBM_REV1-MZ.tex
\documentclass[aoas, preprint]{imsart}

\usepackage{amsthm,amsmath,natbib}
\usepackage{bm}
\usepackage{graphicx}
\usepackage{multirow}
\usepackage{booktabs}
\usepackage{makecell}
\allowdisplaybreaks
\RequirePackage[colorlinks,citecolor=blue,urlcolor=blue]{hyperref}
\startlocaldefs
\endlocaldefs

\begin{document}

\begin{frontmatter}

\title{A Continuous-time Stochastic Block Model for Basketball Networks}


\author{\fnms{Lu} \snm{Xin}\thanksref{m1}\ead[label=e1]{lxin@uwaterloo.ca}}
\address{\printead{e1}}
\author{\fnms{Mu} \snm{Zhu}\thanksref{m1}\ead[label=e2]{m3zhu@uwaterloo.ca}}
\address{\printead{e2}}
\and
\author{\fnms{Hugh} \snm{Chipman}\thanksref{m2}\ead[label=e3]{hugh.chipman@acadiau.ca}}
\address{\printead{e3}}
\affiliation{Department of Statistics and Actuarial Science, University of Waterloo, Waterloo, ON, Canada \thanksmark{m1} and Department of Mathematics and Statistics, Acadia University, Wolfville, NS, Canada \thanksmark{m2}}

\begin{abstract}
For professional basketball, finding valuable and suitable players is the key to building a winning team. To deal with such challenges, basketball managers, scouts and coaches are increasingly turning to analytics. Objective evaluation of players and teams has always been the top goal of basketball analytics. Typical statistical analytics mainly focuses on the box score and has developed various metrics. In spite of the more and more advanced methods, metrics built upon box score statistics provide limited information about how players interact with each other. Two players with similar box scores may deliver distinct team plays. Thus professional basketball scouts have to watch real games to evaluate players. Live scouting is effective, but suffers from inefficiency and subjectivity. In this paper, we go beyond the static box score and model basketball games as dynamic networks. The proposed Continuous-time Stochastic Block Model clusters the players according to their playing style and performance. The model provides cluster-specific estimates of the effectiveness of players at scoring, rebounding, stealing, etc, and also captures player interaction patterns within and between clusters. By clustering similar players together, the model can help basketball scouts to narrow down the search space. Moreover, the model is able to reveal the subtle differences in the offensive strategies of different teams. An application to NBA basketball games illustrates the performance of the model.
\end{abstract}



\end{frontmatter}

\input{Section1}

\input{Section2}

\input{Section3-MZb}

\input{Section4-MZb}

\input{Section5-rev-MZ}

\input{Section6-MZ}

\appendix

\input{Appendix-MZb-rev}

\input{Appendix-CI}

\newpage

\bibliography{Xreference}

\end{document}

%% file: Section1.tex
\section{Introduction}
For decades, basketball data analysis has gained enormous attention from basketball professionals and basketball enthusiasts from various fields. The top goal has always been to better understand how players and teams play, and conduct evaluations more efficiently and objectively. Over the last few years, the explosion of available data, the growth of computer power and the developments of statistical models have made complex modeling of basketball data possible. A revolution is happening in the field of basketball data analysis.

The traditional approaches focus on the box score, which lists the statistics of players and teams of each game, for example, number of field goals attempted, field goals made, rebounds, blocks, steals, plus-minus($+/-$), and other snapshot statistics. By combining the box score statistics, empirically or through regression analysis, various metrics have been developed to evaluate player and team performances \citep{Oliver, Shea}. However, ``there is no Holy Grail of player statistics'' \citep{Oliver}. As pointed out by \citet{Shea}, the metrics are either ``bottom up'' or ``top down''. Bottom-up metrics mostly focus on the individual performance, whereas top-down metrics put emphasis at the team level. Traditional box score metrics mostly fail to take into account two important factors of basketball: the interaction of players and the fact that a basketball play is a real-time process.

Recently, researchers have started to investigate basketball games from these two perspectives. By treating player positions (point guard, shooting guard, small forward, power forward and center) as network nodes and ball passes as network edges, \citet{Bbnet} advocate ``Basketball is not a game, but a network''. They illustrate ball transition patterns of different teams by their basketball networks. Additionally, they quantitatively analyze basketball games and teams by calculating network properties such as degree centrality, clustering coefficient, network entropy and flow centrality. However, when building the networks, \citet{Bbnet} only consider the cumulative passes of games. Hence, the networks are not able to capture details of basketball plays. Neither can they describe players' individual performances. In 2013, the National Basketball Association(NBA) installed optical tracking systems (SportVU technology) in all thirty courts to collect real-time data. The tracking system records the spatial position of the ball and the positions of all players on the court at any time of the game. It also records all actions of the games. Using such comprehensive data, \citet{Bornn2} model the evolution of a basketball play as a complex stochastic process. Their model reveals both offensive and defensive strategies of players and teams. Ultimately, the model estimates the expected scores an offensive team can make at any time of the play. The two approaches above certainly provide more insights and more accurate evaluations of players, teams and basketball plays.

In the NBA, teams obtain new players through trades, free agency and the annual draft. There are so many potential players, especially college players, that no scout is able to keep close track on all of them. Clustering players to a number of groups, according to their performances and playing styles, can efficiently narrow down the target space. When searching for players, basketball managers, scouts and coaches always hope that the new player can quickly fit in the current team. Therefore, how players interact with teammates is of great importance. This must be taken into account during the clustering procedure.

In this paper, we propose a Continuous-time Stochastic Block Model (CSBM) to address the problem of player clustering. We model basketball games as transactional networks and a basketball play as an inhomogeneous continuous-time Markov chain. The CSBM clusters the players according to their performances on the court. It also effectively reveals the players' play styles and the teams' offensive strategies.

The remainder of the paper is organized as follows. In Section 2, we present our view of basketball games as transactional networks and show the data format. In Section 3, we introduce the standard Stochastic Block Model and construct the Continuous-time Stochastic Block Model. An EM algorithm and a complementary algorithm are developed in Section 4. We illustrate our model by an application to NBA basketball games in Section 5. In the end, we summarize our contributions.

%% file: Section2.tex
\section{Basketball Networks} \label{sec:BN}
We begin with a brief introduction to some typical networks, before moving to basketball networks.

\begin{figure}[h]
\begin{center}
\includegraphics[width=100mm]{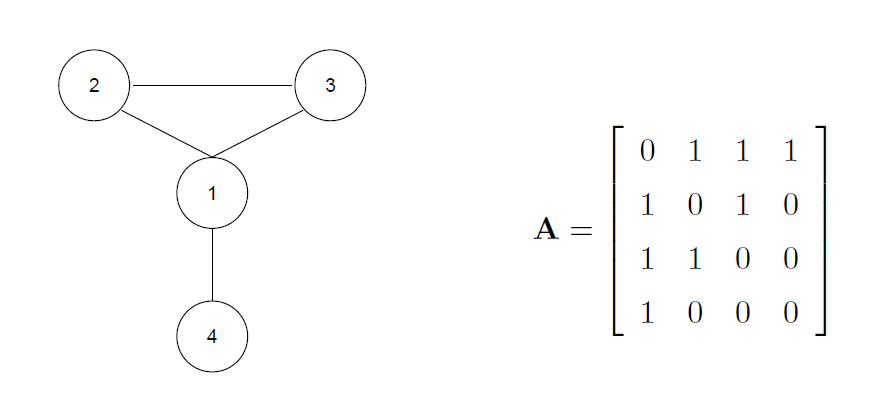}
\end{center}
\caption{An undirected static network (left) and its adjacency matrix (right)} \label{network.example}
\end{figure}

A static network is a graph $G=(V,E)$, consisting of a set of vertices (or nodes) $V$ and a set of edges $E$. A network with $n$ vertices can be represented by an $n\times n$ adjacency matrix, $\mathbf{A}=[A_{ij}]$, where $A_{ij}=0$ or $1$ indicates the absence or presence of the $i\rightarrow j$ edge. Figure~\ref{network.example} shows a simple undirected network ($A_{ij}=A_{ji}$ for all $i$ and $j$) with four vertices and four edges.
The relations between pairs of nodes do not have to be binary-valued. When entries of the adjacency matrix take values other than 0 or 1, the network is called weighted or multi-edged.

Under certain circumstances, instead of observing an edge between two nodes, we observe a series of transactions, for example, phone calls among a number of people in a period of time. Such networks are transactional networks. The corresponding data, as shown in Table \ref{transactional.net}, simply records the senders, the recipients and the time of transactions.
\begin{table}[h]
\caption{A transactional network} \label{transactional.net}
\begin{center}
\begin{tabular}{ccc} \toprule
 \textbf{From} & \textbf{To} & \textbf{Time of transaction} \\
\midrule
1 & 4 & 03/29/2015, 08:27\\

1 & 7 & 03/29/2015, 09:01 \\

3 & 1 & 03/30/2015, 17:11 \\

$\vdots$ & $\vdots$ & $\vdots$ \\
\bottomrule

\end{tabular}
\end{center}
\end{table}

\begin{figure} [h]
\begin{center}
\includegraphics[width=\textwidth]{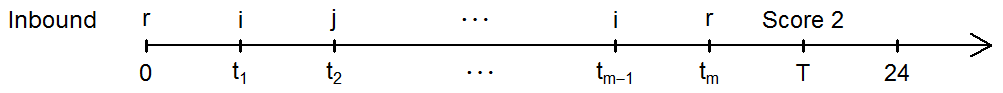}
\caption{A basketball play. The ball is inbounded to player $r$ at time 0; $r$ passes the ball to $i$ at time $t_1$; $i$ passes the ball to $j$ at time $t_2$; $\ldots$; player $i$ receives the ball at time $t_{m-1}$ and passes it to $r$ at time $t_{m}$; the play ends when player $r$ scores 2 points at time $T< 24$ seconds. } \label{BP}
\end{center}
\end{figure}

We now look at basketball, a team game. Players pass the ball to each other and form networks, with players as vertices and passing as transactions on edges. A basketball game is made of basketball plays. Generally, a basketball play starts with inbounding, rebounding, or stealing the ball. During a play, the team with the ball plays offense and the other team plays defense. A play ends when the offensive team shoots the ball (scores or misses but the ball hits the rim), makes a turnover, or the offensive player is fouled when shooting the ball, etc. In the NBA, the time limit for one play is 24 seconds. Figure \ref{BP} illustrates one basketball play.

\begin{figure} [h]
\begin{center}
\includegraphics[width=\textwidth]{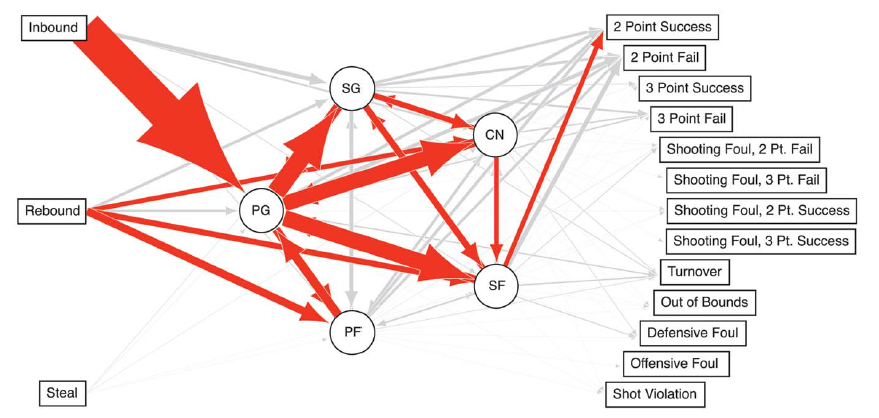}
\caption{Weighted basketball network of 16 NBA games between 16 teams \citep{Bbnet}. Circles represent the five positions (point guard, shooting guard, small forward, power forward, and center), and rectangles represent start or end points of a play. The width of the edge is proportional to the frequency of the corresponding ball transitions. The most frequent transition directions, which sum up to 60\%, are colored red.} \label{bnet}
\end{center}
\end{figure}

In a 48-minute NBA game, a team obtains about 90-110 plays. \citet{Bbnet} model basketball games as weighted networks by counting the frequencies of ball transitions among the starts/ends of plays and the five positions of basketball players (point guard, shooting guard, small forward, power forward and center). Figure \ref{bnet}, which is taken from \citet{Bbnet}, displays the overall weighted network of 16 NBA games between 16 teams they have studied.
The network illustrates play patterns and strategies on a game level. \citet{Bbnet} compare the teams by investigating their networks. However, such network can not capture any detail of real-time basketball play.

We explore basketball at the play level and take into account time effect. More specifically, we regard basketball as a transactional network. Table \ref{tbnet} illustrates our data, from games 1 and 5 of the 2012 NBA eastern conference finals between the Miami Heat and the Boston Celtics. We manually collected the data by watching the videos of the games.
\begin{table}[h]
\caption{Two plays from game 1 of the 2012 NBA eastern conference finals between the Boston Celtics and the Miami Heat. The top three lines show one play for the Boston Celtics. The ball is inbounded to C\#9 (Rajon Rondo) at time 0; Rondo dribbles the ball and passes it to C\#5 (Kevin Garnett) at second 11; Garnett misses a 2-pointer shot at second 12. Lines 4 to 9 illustrate one play for the Miami Heat.} \label{tbnet}

\begin{center}
\begin{tabular}{cccc}  \toprule
 \textbf{From} & \textbf{To} & \textbf{Time(s)} & \textbf{Players on the court}\\
\midrule
\textbf{Inbound} & C\#9 & 0 & C\#9, C\#20, C\#30, C\#34\\

C\#9 & C\#5 & 11 & C\#5, C\#9, C\#20, C\#30, C\#34\\

C\#5 & \textbf{Miss 2} & 12 & C\#5, C\#9, C\#20, C\#30, C\#34\\

\\

\textbf{Rebound} & H\#6 & 0 & H\#3, H\#6, H\#15, H\#21, H\#31 \\

H\#6 & H\#3 & 7 & H\#3, H\#6, H\#15, H\#21, H\#31 \\

H\#3 & H\#15 & 8 & H\#3, H\#6, H\#15, H\#21, H\#31 \\

H\#15 & H\#3 & 9 & H\#3, H\#6, H\#15, H\#21, H\#31 \\

H\#3 & H\#6 & 12 & H\#3, H\#6, H\#15, H\#21, H\#31 \\

H\#6 & \textbf{Miss} 3 & 17 & H\#3, H\#6, H\#15, H\#21, H\#31 \\

\bottomrule
\end{tabular}
\end{center}
\end{table}

In a basketball game, only ten players, five from each team, are on the court at one time. This means a basketball game is subject to many player substitutions. The last column of Table \ref{tbnet} records the players from the offensive team who are on the court at the events. Such information is necessary for our model.
Note that the player inbounding the ball is treated as being off the court at the time of that event.  For example, in Table~\ref{tbnet}, C\#5 is inbounding the ball and not listed as being on the court.

As indicated earlier and shown in Figure \ref{bnet}, there are various ways to start and end a play. A play mostly starts with one of the three initial actions: inbounding, rebounding and stealing the ball. However, a play technically may end with about fifteen different outcomes. For simplicity, we combine the outcomes to six categories: making a 2-pointer (Make 2), making a 3-pointer (Make 3), missing a 2-pointer (Miss 2), missing a 3-pointer (Miss 3), being fouled (Fouled) and making a turnover (TO). Scoring and being fouled at the same time is simply counted as scoring. Catching an air ball is counted as rebounding. All possible ways of giving up the possession of the ball such as direct turnover, being out of bound and offensive foul are regarded as turnover.  We do not consider rare events such as a jump ball. We simply discard the rows corresponding to the rare events.

Although we group events into plays in Table~\ref{tbnet}, the model developed later in Section 3 will treat each event as an individual occurrence, ignoring which play it belongs to.  That is, the data in Table~\ref{tbnet} will be seen as 9 isolated events (each with a timestamp), rather than 3 events in one play and 6 events in another play.

%% file: Section3-MZb.tex
\section{Models}
Our goal is to model the basketball network and cluster players into different groups, so that players in the same group have similar playing styles, while those in different groups play the game in more distinct ways. We propose a Continuous-time Stochastic Block Model. The main idea is to adopt the Stochastic Block Model framework and model basketball plays as Markov Chains.

\subsection{Stochastic Block Models}
The Stochastic Block Model (SBM) \citep{Snijders1997, firststep, wangwong} is an important framework for model-based community detection in static networks. Many recent works have generalized the model \citep{Airoldi2008, Karrer2011} and explored its theoretical properties \citep{Bickel2009, Rohe2011, Zhao2012, Choi2012}. The standard SBM assumes that each node belongs to an underlying block or community. Nodes in the same block are stochastically equivalent. The distribution of an edge between two nodes is governed by the blocks to which they belong. Moreover, given the block affiliations of the nodes, all edges are conditionally independent.

Mathematically, recall that a network with $n$ vertices can be represented by an $n\times n$ adjacency matrix, $\mathbf{A}=[A_{ij}]$, where $A_{ij}=0$ or $1$ respectively indicates the absence or presence of the edge, $i\rightarrow j$. The SBM specifies that, given $K$ blocks and the block labels of all the nodes, $\mathbf{e}=\{e_{1}, e_{2},\ldots,e_{n}\}$, where $e_i\in \{1,2,\ldots,K\}$, the conditional distribution of these $A_{ij}$'s has the form
\begin{equation} \label{condi.like}
\mathcal{L}(\mathbf{A}|\mathbf{e})=\prod_{1\leq i\neq j\leq n}\mathcal{P}(A_{ij}|e_i,e_j),
\end{equation}
where $\mathcal{P}(A_{ij}|e_i,e_j)$ is the conditional probability that there is an edge from $i$ to $j$ given their block labels $e_i$ and $e_j$, typically modeled by a Bernoulli distribution, i.e.,
\begin{equation} \label{ber}
\mathcal{P}(A_{ij}|e_i,e_j)=P_{e_ie_j}^{A_{ij}}(1-P_{e_ie_j})^{1-A_{ij}},
\end{equation}
where $\{P_{kl}:k,l=1,2,\ldots,K\}$ are the $K^2$ parameters of the model.

Given a network and a fixed number of blocks, $K$, the best label configuration $\mathbf{e}$ can be obtained by maximizing the \textit{profile likelihood function} \citep{Bickel2009}. However, finding the optimal solution is NP-hard. Heuristic algorithms are available \citep{Bickel2009, Karrer2011, Zhao2012}. The model can also be fitted with an EM algorithm \citep{Snijders1997}.


\subsection{A Continuous-time Stochastic Block Model} \label{subsec: CSBM.sec}

We generalize the standard SBM to a Continuous-time SBM for basketball networks. During a basketball play (Figure \ref{BP}), an initial action (e.g. inbounding) first transfers the ball to a player; the ball then moves among the players; finally, a play outcome is reached (e.g. the attacking team scores a 2-pointer). Hence, the ball moves among three types of nodes (see Section \ref{sec:BN}): a set of nodes $\mathcal{S}=\{\text{inbounding}, \text{rebounding}, \text{stealing}\}$ that designate different initial states, a total of $n$ nodes that are players themselves, and a set of nodes $\mathcal{A}=\{\text{Make 2}, \text{Miss 2}, \text{Make 3}, \text{Miss 3}, \text{Fouled}, \text{TO}\}$ that designate different outcomes. In addition, we assume that there are $K$ blocks, and each player only belongs to one block. The initial actions and the play outcomes are observable, but the blocks to which the players belong are not. Again, denote the block labels of the players by $\mathbf{e}=\{e_{1}, e_{2},\ldots,e_{n}\}$, where $e_i\in \{1,2,\ldots,K\}$. These block labels are latent. Following the conditional independence assumption of the SBM, the transactions among the nodes are independent given the block labels of the players. The conditional distribution for the entire basketball network, which includes all basketball plays, can be written as:
\begin{multline} \label{conlike.bnet}
\mathcal{L}(\mathbf{T}|\mathbf{e})= \\
\left[\prod_{s\in\mathcal{S}}\prod_{i=1}^n\mathcal{L}^I(\mathbf{T}_{si}|\mathbf{e})\right] \cdot
\left[\prod_{1\leq i\neq j\leq n}\mathcal{L}^P(\mathbf{T}_{ij}|\mathbf{e})\right] \cdot
\left[\prod_{i=1}^{n}\prod_{a\in\mathcal{A}}\mathcal{L}^O(\mathbf{T}_{ia}|\mathbf{e})\right].
\end{multline}
where $\mathbf{T}_{si}$ denotes the transactions from an initial action $s$ to player $i$; $\mathbf{T}_{ij}$ denotes the transactions from player $i$ to player $j$; and $\mathbf{T}_{ia}$ denotes the transactions from player $i$ to an outcome $a$. The conditional distribution \eqref{conlike.bnet} contains three natural components: $\mathcal{L}^I$,  the distribution of all transactions from initial actions to players; $\mathcal{L}^P$, the distribution of all passes among players; and $\mathcal{L}^O$, the distribution of all transactions from players to play outcomes. In the following subsections, we specify the details of these components one by one.

\subsubsection{Transactions from initial actions to players} \label{initial}

Define $\mathbf{P}=\{P_{sk}:s\in\mathcal{S};k=1,2,\ldots,K\}$, where each $P_{sk}$ is the probability that the basketball moves from initial action $s$ to a player in block $k$. These probabilities are subject to the constraint that
\begin{equation}
\sum_{k=1}^KP_{sk}=1,\mbox{ for any } s\in\mathcal{S}.
\end{equation}
Given the block labels of all players, $\mathbf{e}=\{e_1,e_2,...,e_n\}$, the distribution of the transactions from initial action $s$ to players $i$ is defined as
\begin{equation} \label{initiation}
\mathcal{L}^I(\mathbf{T}_{si}|\mathbf{e})=\prod_{h=1}^{m_{si}}\big(P_{se_i}\cdot \frac{1}{G_{e_i}^{sih}}\big),
\end{equation}
where $m_{si}$ is the total number of times that a play goes from initial action $s$ to player $i$. The quantity, $G_{e_i}^{sih}$, denotes the total number of ``eligible receivers'' belonging to block $e_i$ for this particular play (from $s$ to $i$), where ``eligible receivers'' are those players (including $i$ here) who are on $i$'s team and also physically on the basketball court (as opposed to sitting on the bench) at the $h^{th}$ time that a transaction takes place from initial action $s$ to player $i$. In general, we use the notation $G_k^{\triangle}$ to indicate the number of ``eligible receivers'' in block $k$ at the time of an event indexed by $\triangle$. Quantities of this kind will appear a few more times in the next few sections.

The definition \eqref{initiation} implies that players in the same cluster are stochastically equivalent. The probability that player $i$ receives the ball from an initial action $s$ is governed by the block-level probability $P_{se_i}$ and individual-level probability $1/{G_{e_i}^{sih}}$, where we have assumed that all eligible receivers in the same cluster have an equal chance to receive the ball. The individual-level probability is needed in addition to the block-level probability because there is only one ball at all times and only one player can receive it.

Recall that we consider three initial actions: inbounding, rebounding and stealing. While rebounding and stealing both guarantee a new play,  inbounding can start a new play or happen in the middle of a play. For example, a team may call a time-out in the middle of a play, and the play is resumed from the stoppage time by inbounding the ball. Another common situation is when an offensive player is fouled without being awarded free throws, the play is paused and  resumed by inbounding the ball. We treat all inbounding events as initial actions and account for them in this part ($\mathcal{L}^I$) of the probability distribution.

\subsubsection{Transactions among players} \label{passes}

Intuitively, in a basketball play, what happens next mostly depends on the current situation, e.g., who has the ball at the moment, which players are on the court, and so on. Therefore, we model each basketball play as an inhomogeneous Markov chain. Players are treated as regular states; initial actions are treated as initial states; and play outcomes are modeled as absorbing states. We discussed transactions from initial states to regular states in Section~\ref{initial}. In this section, we focus on the regular states and construct $\prod_{1\leq i\neq j\leq n}\mathcal{L}^P(\mathbf{T}_{ij}|\mathbf{e})$ --- the second component in \eqref{conlike.bnet}, the conditional distribution of transactions among players, given the cluster labels $\mathbf{e}$.

\paragraph{Inhomogeneous Poisson process}

\begin{figure} [placement h]
\begin{center}
\includegraphics[width=110mm]{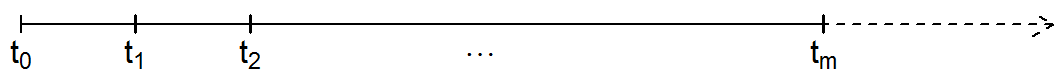}
\caption{A Poisson process} \label{poisson.proc}
\end{center}
\end{figure}

Before doing so, we digress momentarily to look at the distribution of a inhomogeneous Poisson process, often used in event history analysis \citep{Cook.book}. Figure \ref{poisson.proc} shows a Poisson process with $m$ events, happening at times $t_1<\cdots<t_m$ over the interval $[t_0,t_m]$. Suppose that our observation of the process stops at time $t_m$.
Let $\rho(t)$ denote the rate function of this inhomogeneous Poisson process. The distribution is of the form \citep[p.~30]{Cook.book}
\begin{equation} \label{pdf}
\mathcal{L}=\prod_{i=1}^m\mathcal{L}\Big((t_{i-1},t_i]\Big)=\prod_{i=1}^{m}\Big(\rho(t_i)\cdot \exp\big({-\int_{t_{i-1}}^{t_i}\rho(u)\mathrm{d}u}\big)\Big).
\end{equation}
The time intervals $\{(t_{i-1},t_i]$, $i=1,2,\ldots,m\}$ are independent. For each time interval $(t_{i-1},t_i]$, the distribution consists of two parts: the part for the actual event, $\rho(t_i)$, and part for the the time gap between events, $\exp\big({-\int_{t_{i-1}}^{t_i}\rho(u)du}\big)$. The derivation of \eqref{pdf}, especially showing why the part for the time gap has this particular form, is given in Appendix~\ref{appdx:Poisson}; it can also be found in \citet{Cook.book}.

\paragraph{Components of $\mathcal{L}^P(\mathbf{T}_{ij}|\mathbf{e})$}

\begin{figure} [h]
\begin{center}
\includegraphics[width=\textwidth]{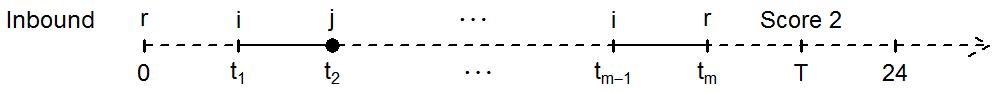}
\caption{Segments of a play that are related to the $i\rightarrow j$ process. The $i\rightarrow j$ process consists of the solid point and the solid segments.} \label{BP2}
\end{center}
\end{figure}

We now derive $\mathcal{L}^P(\mathbf{T}_{ij}|\mathbf{e})$, the conditional distribution of transactions from player $i$ to $j$. To start, we revisit the basketball play shown in Figure \ref{BP} and isolate the segments related to the $i\rightarrow j$ process. For simplicity, suppose that player $j$ is on the court during the entire play. As shown in Figure~\ref{BP2}, player $i$ first receives the ball at time $t_1$ and passes it to player $j$ at time $t_2$, so the time period $(t_1,t_2]$ clearly belongs to the $i\rightarrow j$ process. Next, player $i$ gains possession of the ball again at time $t_{m-1}$ and the ball is passed to player $r \neq j$ at time $t_m$. Although player $i$ does not make this pass to player $j$, he has the \emph{potential} to do so. Hence, the time period $(t_{m-1},t_m)$ is also related to the $i\rightarrow j$ process. In fact, aside from the time point $t_2$ itself, there is no difference between the segments $(t_{m-1},t_m)$ and $(t_1, t_2)$ in terms of being part of the $i\rightarrow j$ process --- as long as $i$ has possession of the ball, the segment is related to the $i\rightarrow j$ process, regardless of whether $i$ actually passes the ball to $j$ or not at the end of the segment. In Figure \ref{BP2}, the segments related to the $i\rightarrow j$ process are highlighted by solid points and segments. Any solid point indicates an actual pass going from $i$ to $j$. Any solid segment means that, during that time period, an $i$-to-$j$ pass has the potential to happen.

Given the cluster labels $\mathbf{e}$, we model each $i\rightarrow j$ process as pieces of a Poisson process. In addition, since each play is independent of one another, we can pool together all the ``solid segments'' and ``solid points'' (again, see Figure~\ref{BP2}) from different plays. Instead of $K^2$ scalar parameters for the standard SBM, for the continuous-time SBM we now have $K^2$ rate functions, $\{\rho_{kl}(t):k,l=1,2,\ldots,K\}$, where each $\rho_{kl}(t)$ is the rate that the ball moves from a player in cluster $k$ to a player in cluster $l$ at time $t$. By equation~\eqref{pdf}, the distribution of transactions from $i$ to $j$ is
\begin{multline} \label{condi.lp}
\mathcal{L}^P(\mathbf{T}_{ij}|\mathbf{e})= \\
\underbrace{%
\left[\prod_{h=1}^{m_{ij}}\Big(\rho_{e_ie_j}(t_{ijh})\cdot \frac{1}{G_{e_j}^{ijh}}\Big)\right]}_{%
\mathcal{L}^{P_1}(\mathbf{T}_{ij}|\mathbf{e})}
\cdot
\underbrace{%
\left[
\prod_{h=1}^{M_i}\exp\Big(-\int_{t_{ih}^-}^{t_{ih}}\rho_{e_ie_j}(t)\cdot \frac{I_j^{ih}}{G_{e_j}^{ih}}\mathrm{d}t\Big)
\right]}_{\widetilde{\mathcal{L}}^{P_2}(\mathbf{T}_{ij}|\mathbf{e})},
\end{multline}
where
\begin{itemize}
\item $m_{ij}$ is the total number of passes from $i$ to $j$;
\item $t_{ijh}$ is the time of the $h^{th}$ pass from $i$ to $j$;
\item $G_{e_j}^{ijh}$ is the number of ``eligible receivers'' belonging to block $e_j$ for the $h^{th}$ pass between $i$ and $j$, with ``eligible receivers''  being those players (excluding $i$ here) who are on $i$'s team and also physically on the basketball court at the time of this pass;
\item $M_i$ is the total number of times that player $i$ has possession of the ball;
\item $(t_{ih}^-, t_{ih})$ is the $h^{th}$ time interval in which player $i$ has possession of the ball;
\item $G_{e_j}^{ih}$ is the number of ``eligible receivers'' belonging to block $e_j$ for the $h^{th}$ pass from player $i$ (regardless of whether $j$ is the recipient or not); and
\item the indicator $I_j^{ih}$ is defined as
\begin{equation}
I_j^{ih} =
\begin{cases} 1, &\mbox{if player $j$ is an ``eligible receiver'' for the $h^{th}$ pass from $i$; } \nonumber \\
0, & \mbox{otherwise.}
\end{cases}
\end{equation}
\end{itemize}
Note that the quantities, $G_{e_j}^{ih}$ and $I_j^{ih}$, are both constant on any interval $(t_{ih}^-, t_{ih}]$, since the rules of the game prevent player substitutions during any such time interval. In addition, we have defined
\begin{equation}
\mathcal{L}^{P_1}(\mathbf{T}_{ij}|\mathbf{e}) \equiv \prod_{h=1}^{m_{ij}}\Big(\rho_{e_ie_j}(t_{ijh})\cdot \frac{1}{G_{e_j}^{ijh}}\Big)
\label{condi.lp1}
\end{equation}
but written $\widetilde{\mathcal{L}}^{P_2}$ for the second component (rather than $\mathcal{L}^{P_2}$) because it can be simplified further (more details below) and this here is not the final expression we shall use.

In \eqref{condi.lp}, the first term contains information about all passes from $i$ to $j$, and the second term contains the information that $i$ does not make a pass to $j$ during all those time gaps in which $i$ has possession of the ball. The overall rate function for the $i\rightarrow j$ process consists of two distinctive parts. First, the rate function $\rho_{e_ie_j}(t)$ captures the rate of passing the ball at a cluster level. Second, similar to the fraction in \eqref{initiation}, the fractions,
\[
\frac{1}{G_{e_j}^{ijh}} \quad\mbox{and}\quad \frac{I_j^{ih}}{G_{e_j}^{ih}},
\]
are the probabilities that player $j$ is the actual receiver of the ball in group $e_j$. As in Section~\ref{initial}, we have assumed that all eligible receivers in the same cluster have an equal chance to receive the ball.

Notice that, if player $j$ is off the court for a particular pass from $i$ or if $j$ is on the opponent team playing against $i$, then the fraction $I_j^{ih}/G_{e_j}^{ih}$ is automatically $0$ by the definition of $I_j^{ih}$. In this way, time intervals $(t_{ih}^-, t_{ih})$ in which $j$ is not an ``eligible receiver'' do not contribute to the $i\rightarrow j$ process, as one intuitively would expect. Furthermore, if $G_{e_j}^{ih}=0$, it means there is no ``eligible receiver'' in block $e_j$ --- this can only happen if player $j$ is not eligible itself, i.e., when $I_j^{ih}=0$, because otherwise $G_{e_j}^{ih}$ is at least one since player $j$ (always) belongs to block $e_j$. We define $0/0=0$. Finally, all time points, $\{t_{ijh}: i,j=1,2,\ldots,n; h=1,2,\ldots,m_{ij}\}$ and $\{t_{ih}^-,t_{ih}: i=1,2,\ldots,n; h=1,2,\ldots,M_i\}$, take values on the interval $[0, 24]$ (see Section \ref{sec:BN}).

\paragraph{Further simplification of $\widetilde{\mathcal{L}}^{P_2}$}

So far, we have derived the (conditional) distribution of transactions from player $i$ to player $j$, $\mathcal{L}^P(\mathbf{T}_{ij}|\mathbf{e})$. The conditional independence assumption means the (conditional) distribution of transactions between all pairs of players is simply
\[
\prod_{1\leq i\neq j\leq n}\mathcal{L}^P(\mathbf{T}_{ij}|\mathbf{e})=
\left[\prod_{1\leq i\neq j\leq n}\mathcal{L}^{P_1}(\mathbf{T}_{ij}|\mathbf{e})\right] \cdot
\left[\prod_{1\leq i\neq j\leq n}\widetilde{\mathcal{L}}^{P_2}(\mathbf{T}_{ij}|\mathbf{e})\right].
\]
The second term above can be simplified further. In particular,
\begin{eqnarray}
\quad \prod_{1\leq i\neq j\leq n}\widetilde{\mathcal{L}}^{P_2}(\mathbf{T}_{ij}|\mathbf{e})
&=&
\prod_{1\leq i\neq j\leq n}\prod_{h=1}^{M_i}
 \exp\Big(-\int_{t_{ih}^-}^{t_{ih}}\rho_{e_ie_j}(t)\cdot \frac{I_j^{ih}}{G_{e_j}^{ih}}\mathrm{d}t\Big) \label{lp2-intermediate} \\
&=&\prod_{i=1}^n\prod_{h=1}^{M_i}\prod_{j\neq i}
 \exp\Big(-\int_{t_{ih}^-}^{t_{ih}}\rho_{e_ie_j}(t)\cdot \frac{I_j^{ih}}{G_{e_j}^{ih}}\mathrm{d}t\Big) \nonumber \\
&=&\prod_{i=1}^n\prod_{h=1}^{M_i}
 \exp\left[-\int_{t_{ih}^-}^{t_{ih}}\sum_{j\neq i}\bigg(\rho_{e_ie_j}(t)
 \cdot \frac{I_j^{ih}}{G_{e_j}^{ih}}\bigg)\mathrm{d}t\right] \nonumber \\
&=&\prod_{i=1}^n\prod_{h=1}^{M_i}
 \exp\left[-\int_{t_{ih}^-}^{t_{ih}}\sum_{l=1}^K\sum_{\substack{j\neq i\\e_j=l}}\bigg(\rho_{e_il}(t)
 \cdot \frac{I_j^{ih}}{G_{l}^{ih}}\bigg)\mathrm{d}t\right]   \nonumber \\
&=&\prod_{i=1}^n\prod_{h=1}^{M_i}
 \exp\left[-\int_{t_{ih}^-}^{t_{ih}}\sum_{l=1}^K \bigg(\rho_{e_il}(t)
 \cdot \sum_{\substack{j\neq i\\e_j=l}}\frac{I_j^{ih}}{G_{l}^{ih}}\bigg)\mathrm{d}t\right].
 \nonumber
\end{eqnarray}
Notice that, on the set $e_j=l$, whenever $G_l^{ih}=0$ (i.e., nobody in block $l$ is an eligible receiver), we must have $I_j^{ih}=0$ as well (i.e., player $j$ cannot be an eligible receiver, either, since $e_j=l$ means player $j$ belongs to block $l$). Therefore,
\[
\sum_{\substack{j\neq i\\e_j=l}} \frac{I_j^{ih}}{G_{l}^{ih}}=I(G_l^{ih}>0).
\]
Continuing with \eqref{lp2-intermediate}, this means
\begin{multline}
\prod_{i \leq 1 \neq j \leq n} \widetilde{\mathcal{L}}^{P_2}(\mathbf{T}_{ij}|\mathbf{e}) = \\
\prod_{i=1}^n
 \underbrace{\prod_{h=1}^{M_i}
 \exp\left[-\int_{t_{ih}^-}^{t_{ih}}\sum_{l=1}^K\Big(\rho_{e_il}(t)\cdot I(G_l^{ih}>0)\Big)\mathrm{d}t\right]}_{%
 \mathcal{L}^{P_2}(\mathbf{T}_i|\mathbf{e})}.  \label{condi.lp2}
\end{multline}

\paragraph{Decomposition of $\mathcal{L}^P(\mathbf{T}_{ij}|\mathbf{e})$}

Putting all the pieces together, the conditional distribution of all transactions among players, given the block labels, is of the form
\begin{equation} \label{condi.lpt}
\prod_{1\leq i\neq j\leq n}\mathcal{L}^P(\mathbf{T}_{ij}|\mathbf{e})=
\left[\prod_{1\leq i\neq j\leq n}\mathcal{L}^{P_1}(\mathbf{T}_{ij}|\mathbf{e})\right] \cdot
\left[\prod_{i=1}^n \mathcal{L}^{P_2}(\mathbf{T}_i|\mathbf{e})\right].
\end{equation}
The first component, $\prod_{i\neq j}\mathcal{L}^{P_1}(\mathbf{T}_{ij}|\mathbf{e})$, contains information about all passes from $i$ to $j$. The second component, $\prod_{i=1}^n \mathcal{L}^{P_2}(\mathbf{T}_i|\mathbf{e})$, contains information about all the time gaps in which player $i$ has possession of the ball --- although, admittedly, denoting all these time gaps here by $\mathbf{T}_i$ is a slight abuse of notation.

In equation~\eqref{condi.lp2}, the indicator $I(G_l^{ih}>0)$ is important for two reasons. First, if node $i$ is the only member in group $l$ or if group $l$ is empty, then it is impossible for $i$ to pass the ball to group $l$, so intuitively the rate function $\rho_{e_il}(t)$ should not contribute any information to this part of the probability distribution. Indeed, in either situation, we have $G_l^{ih}=0$, and this indicator effectively ``annihilates'' the contribution of $\rho_{e_i l}$. Second, we can see from \eqref{condi.lp2} that, overall, player $i$ has a rate of $\sum_{l=1}^K\big(\rho_{e_il}(t)\cdot I(G_l^{ih}>0)\big)$ to pass the ball at time $t$. Given $\rho_{kl}(t)$, when there are fewer groups for player $i$ to pass the ball to, its overall rate of passing the ball is automatically reduced by this indicator, which agrees with our intuition about how basketball games are played.

\subsubsection{Transactions from players to play outcomes} \label{outcomes}

The play outcomes are modeled as absorbing states of the Markov chain. Given a set $\mathcal{A}$ of different play outcomes, we define additional rate functions $\{\eta_{ka}(t):k=1,2,\ldots,K;a\in\mathcal{A}\}$, where $\eta_{ka}(t)$ is the rate that a play goes from group $k$ to absorbing state $a$ at time $t$.

Whenever player $i$ has possession of the ball, there exists a possibility that the ball is ``passed'' to an absorbing state, $a$. Analogous to \eqref{condi.lp}, the distribution of transactions from player $i$ to an absorbing state $a$ can be written as
\begin{equation} \label{condi.absorb}
\mathcal{L}^O(\mathbf{T}_{ia}|\mathbf{e})=
\left[\prod_{h=1}^{m_{ia}}\eta_{e_ia}(t_{iah})\right] \cdot
\left[\prod_{h=1}^{M_i}\exp\Big(-\int_{t_{ih}^-}^{t_{ih}}\eta_{e_ia}(t)\mathrm{d}t\Big)\right],
\end{equation}
where $m_{ia}$ is the total number of times that the ball goes from node $i$ to absorbing state $a$; and $t_{iah}$ is the time of the $h^{th}$ event from $i$ to $a$ --- except that we need no longer multiply the rate function $\eta_{e_i a}(\cdot)$ by an additional individual-level probability (such as $1/G_{e_i}^{iah}$), since there aren't multiple options within an absorbing state as there can be multiple players in a cluster. As in \eqref{condi.lp}, the first term contains information about the event times, and the second term contains the information that player $i$ does not ``cause'' the play to end in absorbing state $a$ while in possession of the ball.

Even though being fouled does not always end a play, we still consider being fouled as an ``outcome'' and take account of all fouls in this part ($\mathcal{L}^{O}$) of the probability distribution.

\subsubsection{A Markov chain}

Here is a brief recapitulation of how we have modeled basketball networks (Sections~\ref{initial}, \ref{passes} and \ref{outcomes}) conditional on the cluster labels of the players. There are three types of nodes in the network: \emph{special} nodes that designate initial actions, \emph{regular} nodes that are players themselves, and \emph{terminal} nodes that designate play outcomes. If we isolate any two regular nodes, or a regular node and a terminal node, transactions between those two nodes have been modelled as an inhomogeneous Poisson process. Each basketball play, however, will consist of a sequence of transactions --- typically starting from a special node, travelling across multiple regular nodes, and ending in a terminal node. Each play is thus an inhomogeneous, continuous-time Markov chain, of which the players are regular states and outcomes are absorbing states.

\subsubsection{Nonparametric modeling of rate functions}

We model the rate functions nonparametrically by cubic B-splines:
\begin{align}\label{spline}
\rho_{kl}(t)&=\sum_{p=1}^{P}e^{\beta_{klp}}B_{p}(t), \mbox{ for } k,l=1,2,\ldots,K, \\
\eta_{ka}(t)&=\sum_{p=1}^{P}e^{\psi_{kap}}B_{p}(t), \mbox{ for }k=1,2,\ldots,K \mbox{ and } a\in\mathcal{A},
\end{align}
where $\{B_{1}(t),B_{2}(t),\ldots,B_{P}(t)\}$ are basis functions; and $\bm{\beta}=\{\beta_{klp}: k,l=1,2,\ldots,K; p=1,2,\ldots,P\}$, $\bm{\psi}=\{\psi_{kap}: k=1,2,\ldots,K; a\in\mathcal{A}; p=1,2,\ldots,P\}$ are coefficients. We use exponentiated coefficients, $e^{\beta_{klp}}$ and $e^{\psi_{kap}}$, to ensure that all rate functions are nonnegative.

\subsection{Related Models}

\citet{Hunter} also adopted event history models to deal with transactional networks, but they did not consider block structures. There are also a number of studies about transactional networks in the framework of SBMs. For example, \citet{Hugh2010} focused on the number of transactions, but did not consider the time factor. \citet{Ho2011}, and \citet{XuHero2014} studied networks at discrete time points and used State Space Models to describe intertemporal dynamics. \citet{UCI} had some ideas similar to ours; they focused on generic transactional networks and parameterized the rate/intensity function using a linear model of various network statistics, but their model could not be applied directly to basketball networks.

%% file: Section4-MZb.tex
\section{An EM$^+$ Algorithm} \label{sec:EMplus}

Since the cluster labels, $\mathbf{e}=(e_1,e_2,\ldots,e_n)$, are unknown, we introduce latent variables and adopt the Expectation-Maximization (EM) algorithm to fit the Continuous-time SBM. Due to the complexity of the model, we have found in our experience that the EM algorithm alone can sometimes be trapped in various local optima. Running the EM algorithm with many random starting points helps, but it is quite inefficient. Instead, we have added a complementary heuristic algorithm to run \emph{after} the EM algorithm. We refer to the complementary algorithm as the ``Plus algorithm'' and call our overall algorithm an ``EM$^+$ algorithm''. Empirically, we have found that the EM$^+$ algorithm often reaches a nice optimal point with fewer starting points than does the EM algorithm itself.

\subsection{EM Algorithm}

Let $\bm{z}_i=(z_{i1},z_{i2},\ldots,z_{iK})$ denote a latent label indicator for node $i$, such that
\begin{equation}
z_{ik} = \begin{cases} 1, &\mbox{if node $i$ belongs to cluster $k$; } \\
0, & \mbox{otherwise. } \end{cases}
\end{equation}
Marginally,
\[
\bm{z}_1, \bm{z}_2, ..., \bm{z}_n \overset{iid}{\sim} \text{multinomial}(1,\bm{\pi}),
\quad\text{where}\quad
\bm{\pi}=(\pi_{1},\ldots,\pi_{K}).
\]
We shall use $\Theta=\{\mathbf{P},\bm{\beta},\bm{\psi},\bm{\pi}\}$ to denote all parameters, and $\mathbf{Z}=\{\bm{z_i}:i=1,2,\ldots,n\}$ to denote all latent indicators. The complete likelihood of the Continuous-time SBM is simply the joint distribution of $(\mathbf{T},\mathbf{Z})$ viewed as a function of $\Theta$. To simplify our notation as well as to make more direct references to the models we described in Section 3, in this section we will often suppress $\Theta$ and still write $\mathcal{L}(\mathbf{T},\mathbf{Z})$ instead of
$\mathcal{L}(\Theta; \mathbf{T},\mathbf{Z})$
for the likelihood function. Hence, the complete likelihood is
\begin{equation} \label{condi.com}
\mathcal{L}(\mathbf{T},\mathbf{Z})=\mathcal{L}(\mathbf{T}|\mathbf{Z})\cdot \mathcal{L}(\mathbf{Z}).
\end{equation}
The conditional likelihood $\mathcal{L}(\mathbf{T}|\mathbf{Z})$ is simply a latent-variable-coded version of $\mathcal{L}(\mathbf{T}|\mathbf{e})$ \eqref{conlike.bnet}, that is,
\begin{align} \label{conlike.bnet.z}
\mathcal{L}&(\mathbf{T}|\mathbf{Z})  \\
=&
\bigg[\prod_{s\in\mathcal{S}}\prod_{i=1}^n\mathcal{L}^I(\mathbf{T}_{si}|\mathbf{Z})\bigg] \cdot
\bigg[\prod_{1\leq i\neq j\leq n}\mathcal{L}^P(\mathbf{T}_{ij}|\mathbf{Z})\bigg] \cdot
\bigg[\prod_{i=1}^{n}\prod_{a\in\mathcal{A}}\mathcal{L}^O(\mathbf{T}_{ia}|\mathbf{Z})\bigg] \nonumber \\
=&
\bigg[\prod_{s\in\mathcal{S}}\prod_{i=1}^n\mathcal{L}^I(\mathbf{T}_{si}|\mathbf{Z})\bigg] \cdot
\bigg[\prod_{1\leq i\neq j\leq n}\mathcal{L}^{P_1}(\mathbf{T}_{ij}|\mathbf{Z}) \cdot
      \prod_{i=1}^n \mathcal{L}^{P_2}(\mathbf{T}_i|\mathbf{Z})\bigg] \nonumber \\
&\qquad \qquad \qquad \qquad \qquad \qquad \qquad \qquad  \qquad \:\: \cdot\bigg[\prod_{i=1}^{n}\prod_{a\in\mathcal{A}}\mathcal{L}^O(\mathbf{T}_{ia}|\mathbf{Z}) \bigg] \nonumber,
\end{align}
where the second step above is due to \eqref{condi.lpt}.
More specifically, the components of \eqref{conlike.bnet.z} are simply latent-variable versions of \eqref{initiation}, \eqref{condi.lp1}, \eqref{condi.lp2} and \eqref{condi.absorb}:
\begin{eqnarray}
\qquad\mathcal{L}^I(\mathbf{T}_{si}|\mathbf{Z})&=&
\prod_{k=1}^K \Big[\prod_{h=1}^{m_{si}}\big(P_{sk}\cdot \frac{1}{G_k^{sih}}\big)\Big]^{z_{ik}}, \label{compo1} \\
\qquad\mathcal{L}^{P_1}(\mathbf{T}_{ij}|\mathbf{Z})&=&
\prod_{k=1}^K\prod_{l=1}^K\bigg[\prod_{h=1}^{m_{ij}}\Big(\rho_{kl}(t_{ijh})\cdot \frac{1}{G_l^{ijh}}\Big)\bigg]^{z_{ik}z_{jl}}, \label{compo2} \\
\qquad\mathcal{L}^{P_2}(\mathbf{T}_i|\mathbf{Z})&=&
\prod_{k=1}^K\bigg[\prod_{h=1}^{M_i}\exp\Big(-\sum_{l=1}^K\int_{t_{ih}^-}^{t_{ih}}\rho_{kl}(t)\cdot I(G_l^{ih}>0)\mathrm{d}t\Big)\bigg]^{z_{ik}}, \label{compo3} \\
\qquad\mathcal{L}^O(\mathbf{T}_{ia}|\mathbf{Z})&=&
\prod_{k=1}^K\bigg[\prod_{h=1}^{m_{ia}}\eta_{ka}(t_{iah})\cdot \prod_{h=1}^{M_i}\exp\Big(-\int_{t_{ih}^-}^{t_{ih}}\eta_{ka}(t)\mathrm{d}t\Big)\bigg]^{z_{ik}}. \label{compo4}
\end{eqnarray}
The marginal likelihood of $\mathbf{Z}$ is
\begin{equation} \label{compo5}
\mathcal{L}(\mathbf{Z})=\prod_{i=1}^n\prod_{k=1}^K\pi_k^{z_{ik}}.
\end{equation}

\subsubsection{E-step}

In the E-step, we compute $\mathbf{E}\big(\log\mathcal{L}(\mathbf{T,Z})|\mathbf{T};\Theta^*\big)$, the conditional expectation of the log-likelihood given the observed network $\mathbf{T}$ under the current parameter estimates (denoted by $\Theta^*$). The conditional expectation is with respect to the latent variables $\mathbf{Z}$. From \eqref{compo1}-\eqref{compo4} it is clear (details in the Appendix~\ref{appdx:condEdetail}) that there are three types of conditional expectations to evaluate:
\begin{itemize}
\item $\mathbf{E}(z_{ik}|\mathbf{T};\Theta^*)$, from $\log\mathcal{L}^I(\mathbf{T}_{si}|\mathbf{Z})$, $\log\mathcal{L}^O(\mathbf{T}_{ia}|\mathbf{Z})$ and $\log\mathcal{L}(\mathbf{Z})$, respectively;
\item $\mathbf{E}(z_{ik}z_{jl}|\mathbf{T};\Theta^*)$, from $\log\mathcal{L}^{P_1}(\mathbf{T}_{ij}|\mathbf{Z})$; and
\item $\mathbf{E}\big(z_{ik}\cdot I(G_{l}^{ih}>0)|\mathbf{T};\Theta^*\big)$,
from $\log{\mathcal{L}^{P_2}(\mathbf{T}_i|\mathbf{Z})}$.
\end{itemize}
After taking logarithms, the terms involving $1/G_k^{sih}$ and $1/G_l^{ijh}$ in \eqref{compo1} and \eqref{compo2} are additive ``constants'' that depend only on the latent variables $\mathbf{Z}$ but contain no information about the parameters $\Theta$; they can be omitted for the EM algorithm. The quantity $$G_l^{ih}=\sum_{j\neq i} \big( z_{jl} \cdot I_j^{ih} \big)$$ and hence the indicator $I(G_{l}^{ih}>0)$ are both functions of the latent variables. Here, we see more clearly why the further simplification of $\mathcal{L}^{P_2}$ --- equation~\eqref{condi.lp2} --- is useful. Due to the interactions of the players, the latent variables are conditionally dependent and an exact calculation of the conditional expectations above is NP-hard. For instance, in order to calculate $\mathbf{E}(z_{ik}|\mathbf{T};\Theta^*)$, one needs to marginalize the cluster labels over all nodes that interact with $i$. We use a Gibbs sampler to draw samples from $\mathcal{L}(\mathbf{Z}|\mathbf{T};\Theta^*)$, and use the corresponding sample means to approximate $\mathbf{E}(z_{ik}|\mathbf{T};\Theta^*)$, \linebreak $\mathbf{E}(z_{ik}z_{jl}|\mathbf{T};\Theta^*)$ and $\mathbf{E}\big(z_{ik}\cdot I(G_{l}^{ih}>0)|\mathbf{T};\Theta^*\big)$.

\paragraph{Gibbs sampler}

Let $\mathbf{Z}^{-i}=\{\bm{z_j}:j\neq i\}$ denote the latent cluster indicators of all players other than $i$.
The idea of the Gibbs sampler is to draw
\begin{eqnarray*}
\mathbf{z}_1 &\sim& \mathcal{L}(\mathbf{z}_1|\mathbf{Z}^{-1},\mathbf{T};\Theta^*), \\
\mathbf{z}_2 &\sim& \mathcal{L}(\mathbf{z}_2|\mathbf{Z}^{-2},\mathbf{T};\Theta^*), \\
& & \vdots \\
\mathbf{z}_n &\sim& \mathcal{L}(\mathbf{z}_n|\mathbf{Z}^{-n},\mathbf{T};\Theta^*), \\
\mathbf{z}_1 &\sim& \mathcal{L}(\mathbf{z}_1|\mathbf{Z}^{-1},\mathbf{T};\Theta^*), \\
\mathbf{z}_2 &\sim& \mathcal{L}(\mathbf{z}_2|\mathbf{Z}^{-2},\mathbf{T};\Theta^*), \\
& & \vdots
\end{eqnarray*}
repeatedly until the stationary distribution is reached. (In our application, a handful of repetitions are often sufficient.)
Under the current parameter estimate $\Theta^*$, the conditional distribution of $\bm{z_i}$ given $\mathbf{Z}^{-i}$ and $\mathbf{T}$ is
\begin{equation} \label{sampler}
\mathcal{L}(\bm{z_i}|\mathbf{Z}^{-i},\mathbf{T};\Theta^*)=\frac{\mathcal{L}(\mathbf{T,Z};\Theta^*)}{\sum_{\bm{z_i}}\mathcal{L}(\mathbf{T,Z};\Theta^*)},
\end{equation}
a multinomial distribution which is easy to sample from. More explicitly, suppose that, at the current step, $z_{jc_j}=1$ for $j\neq i$ --- this means $e_j=c_j$ for all $j\neq i$ or that $c_j$ is the current group label for player $j$. Then, the conditional probability of player $i$ belonging to cluster $k$ is
\begin{align}
\mathcal{P}(z_{ik}=1&|\mathbf{Z}^{-i},\mathbf{T};\Theta^*) \label{gibbs} \\
&=\mathcal{P}(e_i=k|\{e_j=c_j:j\neq i\},\mathbf{T};\Theta^*) \nonumber \\
&=\frac{\mathcal{L}(\mathbf{T},\mathbf{e}=(c_1,c_2,\ldots,c_{i-1},k,c_{i+1},\ldots,c_n);\Theta^*)}
{\sum_{l=1}^K\mathcal{L}(\mathbf{T},\mathbf{e}=(c_1,c_2,\ldots,c_{i-1},l,c_{i+1},\ldots,c_n);\Theta^*)}.\nonumber
\end{align}

\subsubsection{M-step}

In the M-step, we update the parameters $\Theta$ by maximizing $\mathbf{E}\big(\log\mathcal{L}(\mathbf{T,Z})|\mathbf{T};\Theta^*\big)$. We have closed-form solutions for $\bm{\pi}$, the marginal probabilities of $\mathbf{Z}$, and for $\mathbf{P}$, the transition probabilities from initial states:
\begin{align}
\pi_{k}&=\frac{\sum_{i=1}^n\mathbf{E}(z_{ik}|\mathbf{T};\Theta^*)}
{\sum_{i=1}^n\sum_{l=1}^{K}\mathbf{E}(z_{il}|\mathbf{T};\Theta^*)}
=\frac{\sum_{i=1}^n\mathbf{E}(z_{ik}|\mathbf{T};\Theta^*)}{n},\label{update.pi} \\
P_{sk}&=\frac{\sum_{i=1}^n[m_{si}\mathbf{E}(z_{ik}|\mathbf{T};\Theta^*)]}
{\sum_{k=1}^K\sum_{i=1}^n[m_{si}\mathbf{E}(z_{ik}|\mathbf{T};\Theta^*)]},  \label{update.iprob}
\end{align}
for $k=1,2,\ldots,K$ and $s\in\mathcal{S}$; detailed derivations are given in Appendix~\ref{appdx:prob-updates}. However, there are no closed-form solutions for $\bm{\beta}$ and $\bm{\psi}$, the (log)-coefficients for the rate functions. We use the quasi-Newton method with L-BFGS-B updates --- more specifically, we use the \verb!optim! function in R and supply with it the analytic form of the gradient.

\subsubsection{Remarks} \label{subsubsec:remarks}

Here, we make a few important remarks about the EM algorithm. The conditional probabilities driving the Gibbs sampler turn out to be fairly close to 0 or 1, that is, in equation \eqref{gibbs}, one of the $K$ terms being summed in the denominator is significantly larger than the others. The reason is that each player is involved in many transactions. As far as the likelihood function is concerned, these transactions act as if they were repeated measurements, which reinforce the assignment of the player to a particular group. The Gibbs sampler thus converges very quickly to a singular probability mass. This essentially reduces the EM algorithm to a $K$-means algorithm: the E-step re-assigns the players to different groups, and the M-step re-estimates the parameters. Overall, the EM algorithm converges in just a few iterations. But the EM algorithm can sometimes be trapped in a local optimum. The typical way to avoid these traps is to use different starting points, run the EM algorithm for a few times, and pick the one giving the largest likelihood value. This ``standard'' procedure alone could be quite inefficient. Instead, we introduce another heuristic algorithm, which we refer to as the Plus algorithm (Section~\ref{sec:plus}), as a complement to the EM algorithm. Sometimes, e.g., when the EM solution is already quite good, the Plus algorithm may not find any further improvement.

\subsection{The Plus Algorithm}
\label{sec:plus}

This algorithm is inspired by the heuristic algorithm used by \citet{Karrer2011} for the so-called degree-corrected SBM. The main idea is to evaluate all neighbors of the current labelling configuration and move to the best neighbor no matter if the likelihood improves or not. A neighbor of a labelling configuration $\mathbf{e}=(e_1,e_2,\ldots,e_n)$ is defined as the one with only one entry being different. Thus, if $\mathbf{e}^{\prime}$ and $\mathbf{e}$ are neighbors, then there exists some $1\leq i\leq n$ such that $e_i\neq e^{\prime}_i$, but otherwise $e_j = e^{\prime}_j$ for all $j \neq i$. Given $n$ nodes and $K$ clusters, one labelling configuration has $n(K-1)$ neighbors. The steps of the algorithm are as follows.

\begin{enumerate}
\item Start with $r=0$.
\item Repeat the following steps until convergence, or for a fixed number of steps.
\begin{enumerate}
\item \label{plus-check-step} Given a labelling configuration $\mathbf{e}^{(r)}$ and parameter $\Theta^{(r)}$ estimated under $\mathbf{e}^{(r)}$, calculate the likelihood of all neighboring configurations, using the same parameter estimate, $\Theta^{(r)}$.
\item\label{plus-update-step} Let $\mathbf{e}^{(r+1)}$ be the neighbor that gives the largest likelihood.
\item\label{plus-para-step} Re-estimate the parameters using $\mathbf{e}^{(r+1)}$, and denote the result by $\Theta^{(r+1)}$.
\end{enumerate}
\item Choose the best configuration among $\mathbf{e}^{(0)}, \mathbf{e}^{(1)}, \mathbf{e}^{(2)}, ...$.
\end{enumerate}

We use the result from the EM algorithm as the starting point $\mathbf{e}^{(0)}$ to run the Plus algorithm. The Plus algorithm converges when there exists a set of configurations $\mathbf{e}_1, \mathbf{e}_2, ..., \mathbf{e}_q$ such that $\mathbf{e}_1$ is the best neighbor of $\mathbf{e}_2$, $\mathbf{e}_2$ is the best neighbor of $\mathbf{e}_3$, ..., and $\mathbf{e}_q$ is the best neighbor of $\mathbf{e}_1$. Often, this happens for $q=2$, but sometimes it can happen for $q>2$. Note that, while $\mathbf{e}^{(r+1)}$ in step (\ref{plus-update-step}) gives the largest likelihood among all neighbors of $\mathbf{e}^{(r)}$, it may still give a smaller likelihood than does $\mathbf{e}^{(r)}$ itself, but the Plus algorithm ``accepts'' $\mathbf{e}^{(r+1)}$ nonetheless. This is the main reason why the Plus algorithm can help the EM algorithm avoid local optima. On the other hand, the Plus algorithm itself moves very slowly --- in any given iteration, only one node label is changed, so it is quite inefficient to use it as a standalone algorithm, but we have found it to work well as a complement to the EM algorithm.

%% file: Section5-rev-MZ.tex
\section{Application to NBA data}
In this section, we apply our Continuous-time Stochastic Block Model (CSBM) to a few NBA basketball games that we have annotated ourselves. The games are: the 2012 NBA eastern conference finals between the Miami Heat and the Boston Celtics, games 1 and 5; and the 2015 NBA finals between the Cleveland Cavaliers and the Golden State Warriors, games 2 and 5. For each game, we only consider the first three quarters to avoid having to deal with garbage time or irregular playing strategies (such as committing fouls on purpose), which are both common in the last quarter. In Section \ref{sec.sa}, we present some further model simplifications and corresponding adjustments to the EM$^+$ algorithm. In Sections \ref{sec.HC} and \ref{sec.GC}, we present results for 2012 games between the Heat and the Celtics, and those for the 2015 games between the Cavaliers and the Warriors, respectively. In Section \ref{sec.HCa}, we compare the 2012 Miami Heat with the 2015 Cleveland Cavaliers, while paying special attention to the performance of LeBron James as he played with these two different teams in those two series.

\subsection{Model simplifications and adjustments of the EM$^+$ algorithm} \label{sec.sa}
In practice, the general model is complex, with $K(K+|\mathcal{A}|)$ rate functions to estimate.
For applications to NBA data, we further
simplify the general form by defining
\begin{align}
\rho_{kl}(t)&= \lambda_{k}(t)\cdot P_{kl} \label{sim1},\\
\eta_{ka}(t)&= \lambda_{k}(t)\cdot P_{ka} \label{sim2},
\end{align}
such that $\lambda_k(t)$ is the rate function of the ball leaving a player in group $k$; $P_{kl}$ and $P_{ka}$ are transition probabilities that the ball goes to group $l$ and absorbing state $a$, respectively. The transition probabilities are subject to the constraint
\begin{equation} \label{const}
\sum_{l=1}^KP_{kl}+\sum_{a\in\mathcal{A}}P_{ka}=1, \mbox{for any } k=1,2,\ldots,K .
\end{equation}
By making such simplifications, we assume that, whenever the ball leaves cluster $k$, the rates to other clusters and absorbing states are formed by a common rate and proportionality constants. In reality, the transition probabilities may change over time, but we believe that the simplified model still contains sufficient information to cluster players and reveal important patterns. The results in next section provide convincing evidence.

The rate function simplifications lead to modifications in the EM$^+$ algorithm. Recall that for the general model, we update the marginal and initial probabilities by \eqref{update.pi} and \eqref{update.iprob}, respectively, in the M-step. Meanwhile, we update the rate functions by the quasi-Newton method. Under the simplified model, \eqref{update.pi} and \eqref{update.iprob} still apply because the marginal and initial probabilities remain unchanged. Nevertheless, the $K(K+|\mathcal{A}|)$ rate functions reduce to $K$ rate functions and a $K\times (K+|\mathcal{A}|)$ transition matrix. We still adopt quasi-Newton for the rate functions, yet we have closed-form solutions for the transition probabilities (details in Appendix~\ref{appdx:condEdetail-simp}),
\begin{equation} \label{update.pkl}
P_{kl}=\frac{\sum_{1\leq i\neq j\leq n}\Big(\mathbf{E}[z_{ik}z_{jl}|\mathbf{T};\Theta^*]\cdot m_{ij}\Big)}{\sum_{i=1}^n \sum_{h=1}^{M_i}\Big(\mathbf{E}\Big[z_{ik}I(G_l^{-i}(t_{ih})>0)\Big|\mathbf{T};\Theta^* \Big]\cdot \int_{t_{ih}^-}^{t_{ih}}\lambda_{k}(t) \mathrm{d}t\Big)+\zeta_k},
\end{equation}
\begin{equation} \label{update.pka}
P_{ka}=\frac{\sum_{i=1}^n\Big(\mathbf{E}[z_{ik}|\mathbf{T};\Theta^*]\cdot m_{ia}\Big)}{\sum_{i=1}^n \sum_{h=1}^{M_i}\Big(\mathbf{E}[z_{ik}|\mathbf{T};\Theta^*]\cdot \int_{t_{ih}^-}^{t_{ih}}\lambda_{k}(t) \mathrm{d}t\Big) +\zeta_k},
\end{equation}
for $k,l=1,2,\ldots,K$ and $a\in\mathcal{A}$. The parameter $\zeta_k$ is the Lagrange multiplier, which can be easily solved by finding the root of
$\sum_{l=1}^KP_{kl}+\sum_{a\in\mathcal{A}}P_{ka}=1$ with the R function \verb!uniroot!.

For this simplified model, all probability parameters including marginal probabilities $\pi_k$, initial probabilities $P_{sk}$ and transition probabilities $P_{kl}$ and $P_{ka}$ have closed-from updates. Hence, to make the EM$^+$ algorithm more efficient, we partition the parameter set $\Theta$ into two groups: $\Theta_{fast} = \{ \pi_k, P_{sk}, P_{kl}, P_{ka} \}$, consisting of all parameters with closed-form updates, and $\Theta_{slow} = \{ \lambda_{k}(t) \}$, consisting of all parameters that we must update with quasi-Newton. Instead of updating all $\Theta$ only in the M-step of the EM algorithm and Step (\ref{plus-para-step}) of the Plus algorithm, parameters belonging to $\Theta_{fast}$ are always updated instantaneously ``on the fly'' --- meaning that they are updated whenever there is a change in $\mathbf{Z}$ or the cluster labels $\mathbf{e}$. More specifically, $\Theta_{fast}$ are updated when calculating each likelihood function in the Gibbs sampler \eqref{gibbs} of the EM algorithm and Step (\ref{plus-check-step}) of the plus algorithm.




\subsection{Miami Heat versus Boston Celtics in 2012} \label{sec.HC}
In the 2012 NBA eastern conference finals, eleven players from the Heat and ten players from the Celtics played in the first three quarters of their 1st and 5th games. We omit two Celtics players, Ryan Hollins and Marquis Daniels, because they each touched the ball only once in those quarters. The data, which have been illustrated in Table \ref{tbnet}, consist of 283 plays (142 for the Heat and 141 for the Celtics) and 1205 transactions (657 for the Heat and 548 for the Celtics). We fit three different CSBMs --- one to the Heat's transactions alone, one to the Celtics' transactions alone, and another one to transactions from both teams pooled together. In what follows, we discuss in detail our clustering results, initial probability estimates, fitted rate functions, and transition probability estimates. Given our data size (11 Heat players and 8 Celtics players), we picked a moderate number of clusters ($K=3$). In practice, since the main purpose of our model is to cluster players and narrow down the search space for basketball scouts, the choice of $K$ will mostly depend on the size of the basketball network and how elaborate one wants the clustering results to be.

\paragraph{Clustering results} The cluster labels for the players are reported in Table \ref{cluster}.
Recall that basketball players play in five different positions: point guard (PG), shooting guard (SG), small forward (SF), power forward (PF) and center (C). Generally speaking, the heights of the players are PG$<$SG$<$SF$<$PF$<$C.

\begin{table}[h]
\caption{Clustering results for the 2011-2012 Miami Heat and Boston Celtics ($K=3$).  Cluster labels are C1, C2, C3.  Three different clustering results are presented (two under ``Alone'' and one under ``Together''). Player positions are included for reference only; they are not used by the clustering algorithm.} \label{cluster}
\begin{center}
\begin{tabular}{|c|l|c|c|c|c|c|c|c|} \toprule
& & & \multicolumn{3}{ c| }{Alone} & \multicolumn{3}{ c| }{Together} \\
\midrule
Team & \multicolumn{1}{ c| }{Player} & Position & C1 & C2 & C3 & C1 & C2 & C3\\
\midrule
\multirow{11}{*}{Heat} & Mario Chalmers & PG & X & & & X & & \\
 & Norris Cole & PG & X & & & X & & \\
 & Dwyane Wade & SG &  &X & &  &X & \\
 & LeBron James & SF &  &X & &  &X & \\
 & James Jones & SG &  & &X &  & &X \\
 & Shane Battier & SF &  & &X &  & &X \\
 & Mike Miller & SF &  & &X &  & &X \\
 & Chris Bosh & PF &  & &X &  & &X \\
 & Udonis Haslem & PF &  & &X &  & &X \\
 & Ronny Turiaf & C &  & &X &  & &X \\
 & Joel Anthony & C &  & &X &  & &X \\ [0.3cm]
 \cline{1-6}
  &&&&&&&&\\
\multirow{8}{*}{Celtics} & Rajon Rondo & PG & X & & & X & & \\
& Keyon Dooling & PG & X & & & X & & \\
& Ray Allen & SG &  &X& &  &X & \\
& Paul Pierce & SF & &X & &  &X & \\
& Mickael Pietrus & SF & & &X &  & &X \\
& Brandon Bass  & PF & & &X &  &X & \\
& Kevin Garnett & C & & &X &  & &X \\
& Greg Stiemsma & C & & &X &  & &X \\[0.3cm]
\bottomrule
\end{tabular}
\end{center}
\end{table}

Considered separately, players in the two teams are clustered in similar manners. Point guards are in cluster 1; two perimeter players --- \{Wade, James\} from the Heat and \{Allen, Pierce\} from the Celtics --- are in cluster 2; and the other players are in cluster 3. Roughly speaking, players with similar heights and close positions are clustered into the same group. Point guards certainly play in a different style than those of power forwards and centers. Shooting guards and small forwards are both perimeter players and often play in similar styles. In our case, Wade, James, Allen and Pierce are different than the other perimeter players, because they are stars. They have extraordinary offensive skills, so they can carry the ball longer and shoot more often.  By contrast, the shooting guards and small forwards in cluster 3 play without the ball for most of the time.

When the two teams are pooled together, only one player (Brandon Bass from the Celtics) switches from cluster 3 to cluster 2. Actually, he is a ``mini'' PF, who has a typical PF's weight and strength but the height of an SF, so his playing style is in between those of a typical SF and a typical PF. When compared only with other Celtics players, he is more similar to those in cluster 3. However, when players from the Heat also are included in the comparison, he starts to look more similar to LeBron James (a strong SF) and very different than those in cluster 3 who are on the Heat, e.g., in terms of rebounding, cutting, post playing, so he is re-clustered into cluster 2.

In our subjective assessment, players in cluster 1 tend to dribble the ball a lot but do not shoot very often, those in cluster 2 both carry and shoot the ball, whereas those in cluster 3 are mostly responsible for catching rebounds and shooting, but not so much for carrying the ball. In what follows, we will see these differences of the three clusters reflected in the different parameters of the CSBM.

\paragraph{Initial probabilities} Table \ref{ini.k3} displays the estimated transition probabilities from each initial action to the three clusters.  Most inbounds go to point guards, because they usually are the ones to carry the ball from the back court to the front court. The Heat inbound more often to cluster 2 than the Celtics do, because LeBron James (in cluster 2) sometimes plays like a point guard. More than half of the rebounds are caught by cluster 3, the tall players. For the Celtics, their cluster 1 players catch almost as many rebounds as those in their cluster 2, because the starting point guard, Rajon Rondo (in cluster 1), is an excellent rebounder. Regarding steals (a relatively rare event), the three clusters contribute equally within the Heat but somewhat differently within the Celtics.
\begin{table}[h]
\caption{Estimated transition probabilities ($P_{sk}$) from each initial action to clusters C1, C2, C3, for three different clustering models of the 2011-2012 Miami Heat and Boston Celtics.} \label{ini.k3}
\begin{tabular}{c|l|ccc} \toprule
 & & C1 & C2 & C3\\
\midrule
\multirow{3}{*}{Heat} &Inbound & 0.716 & 0.194 & 0.090\\

& Rebound & 0.109 & 0.375 & 0.516\\

& Steal & 0.333 & 0.333 & 0.333\\
\midrule
\multirow{3}{*}{Celtics} &Inbound & 0.868 & 0.059 & 0.073\\

& Rebound & 0.188 & 0.208 & 0.604\\

& Steal & 0.375 & 0.500 & 0.125\\
\midrule
\multirow{3}{*}{Together} &Inbound & 0.793 & 0.133 & 0.074\\

& Rebound & 0.143 & 0.357 & 0.500\\

& Steal & 0.364 & 0.454 & 0.182\\
\bottomrule
\end{tabular}
\end{table}

\begin{figure} [h]
\begin{center}
\includegraphics[width=\textwidth]{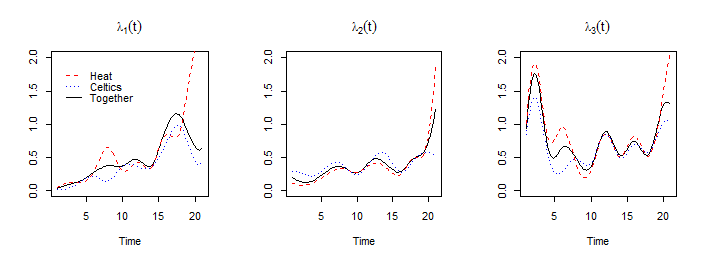}
\caption{Fitted rate functions for the 2011-2012 Miami Heat and Boston Celtics, $\lambda_1(t)$, $\lambda_2(t)$ and $\lambda_3(t)$, each describing the rate with which the ball leaves a player in cluster 1, cluster 2 and cluster 3, respectively.} \label{Rate.k3}
\end{center}
\end{figure}

\paragraph{Rate functions} Figure~\ref{Rate.k3} contains the fitted rate functions $\{\lambda_k(t): k=1, 2, 3\}$ for the ball leaving a player in group $k$. Overall, these functions are quite different for the three clusters. For the same cluster, the rate functions from different teams are similar in general, but have considerable differences at certain time points. Below, we compare the patterns of the rate functions over four distinct time periods: $t\in(0,5)$, $t\in(5,10)$, $t\in(10,15)$, and $t>15$.

At the beginning of a play, it usually takes about five seconds for a point guard to dribble the ball from the back court to the front court. Players in cluster 2 sometimes do that instead of point guards. Therefore, $\lambda_1(t)$ and $\lambda_2(t)$ are low for $t\in(0,5)$. However, for both teams their $\lambda_3(t)$ has a high and sharp peak around $t\approx2$, because players in cluster 3 often catch rebounds and start new plays by quickly passing the ball to those in the other two clusters.

After the ball arrives at the front court, the players spend about 5 seconds to settle down to their offensive layout. During this time period, i.e., $t\in(5,10)$, the two teams have different strategies. For the Heat, the point guards usually pass the ball to either James or Wade and let them handle the ball, so we can see a small peak in the Heat's $\lambda_1(t)$ function. For the Celtics, their point guards --- especially Rondo --- usually continue to hold the ball and organize the offense, so the Celtics' $\lambda_1(t)$ function even declines a little right after $t>5$. The two teams' $\lambda_3(t)$ functions exhibit significant difference over this time period. For the Heat, their players in cluster 3 mostly play as transit ports, i.e., they get the ball and pass it out soon. For the Celtics, their players in cluster 3 --- especially Kevin Garnett --- have more opportunities to handle the ball. That is why in the right panel, the Heat's $\lambda_3(t)$ function has a peak around $t\approx7$, while the Celtics' $\lambda_3(t)$ has a local minimum between $5<t<6$.

For $t \in (10,15)$, if the play still keeps going, players start to pass the ball more frequently and seek scoring opportunities. This is indicated by higher values in $\lambda_1(t)$ and $\lambda_2(t)$ as well as a local peak in $\lambda_3(t)$ on $t\in(10,15)$. During this time period, both teams play in a similar style, and their rate functions almost overlap.

Due to the 24-second time limit for each play, both team increase their offensive pace after $t>15$. However, when time reaches about $t\approx20$, the two teams start to show highly distinctive playing patterns. For the Heat, all three of their rate functions rise rapidly, which means that all of their players tend to release the ball quickly, either passing it on to others or shooting. For the Celtics, their $\lambda_2(t)$ and $\lambda_3(t)$ also rise, but not as much as those of the Heat. The Celtics appear to play with more patience. An unusual phenomenon is that, for the Celtics, their rate function $\lambda_1(t)$ actually decreases after $t>17$. This is because the starting point guard, Rajon Rondo (in cluster 1), is not the best jump shooter. Close to the end of the time limit and against the tough defense from the Heat, he typically struggles a bit trying to pass or shoot, so the ball stays in his hands for a little longer.

In Appendix~\ref{appdxCI}, we provide an expanded version of Figure~\ref{Rate.k3} which includes 95\% confidence bands for these
rate functions.

\paragraph{Transition probabilities} The estimated transition probabilities for events originating from the three different clusters are presented in Table \ref{tran.k3}. We will focus on the transition probabilities of each team alone. When the two teams are pooled together, the estimated transition probabilities simply appear to be averages of the individual team results.

\begin{table}[h]
\caption{Estimated transition probabilities ($P_{kl}$ and $P_{ka}$) for the 2011-2012 Miami Heat and Boston Celtics ($K=3)$.  Rows are originating clusters and columns are receiving clusters and play outcomes.} \label{tran.k3}
\begin{center}
\scriptsize
\begin{tabular}{c|c|ccccccccc} \toprule
 && C1 & C2 & C3 & Make2 & Miss2 & Make3 & Miss3 & Fouled & TO\\
\midrule
\multirow{3}{*}{Heat} & C1 & 0 & 0.564 & 0.296  & 0.035 & 0.014 & 0 & 0.042 & 0.021 & 0.028\\

 & C2 & 0.188 & 0.262 & 0.225 & 0.103  & 0.087& 0.008 & 0.032 & 0.063 & 0.032\\

& C3 & 0.226 & 0.426 & 0.090 & 0.052  & 0.064 & 0.039 & 0.064 & 0.013 & 0.026\\
\midrule
\multirow{3}{*}{Celtics} & C1 & 0.175 & 0.332 & 0.327 & 0.031  & 0.083 & 0.010 & 0.016 & 0.005 & 0.021\\

 & C2 & 0.270 & 0.066 & 0.262 & 0.065  & 0.172 & 0.033 & 0.066 & 0.041 & 0.025\\

& C3 & 0.304 & 0.177 & 0.094 & 0.191  & 0.149 & 0 & 0.014 & 0.057 & 0.014\\
\midrule
\multirow{3}{*}{Together} & C1 & 0.119 & 0.479 & 0.250 & 0.032  & 0.053 & 0.006 & 0.026 & 0.012 & 0.023\\

 & C2 & 0.220 & 0.210 & 0.211 & 0.097  & 0.124 & 0.014 & 0.039 & 0.056 & 0.029\\

& C3 & 0.262 & 0.341 & 0.083 & 0.110  & 0.087 & 0.023 & 0.045 & 0.030 & 0.019\\
\bottomrule
\end{tabular}
\end{center}
\end{table}
First, we look at passes among clusters. For the Heat, James and Wade (both in cluster 2) are the absolute key players for the team, so players from both cluster 1 and cluster 3 tend to pass the ball to them (cluster 2) with very high probabilities (56.4$\%$ and 42.6$\%$, respectively). James and Wade also pass the ball more often to each other than to the other clusters ($26.2\%$ vs. $18.8\%$ and $22.5\%$, respectively). The two players in cluster 1, Chalmers and Cole, do not pass to each other in our data because they are never on the court at the same time during those games. The Celtics, on the other hand, tend to move the ball more evenly among the three clusters. Their clusters 1 and 2 each has almost equal probabilities to pass the ball to the other two clusters. Their transition probabilities are lower within each cluster than between different clusters.

Next, we discuss shooting choices. For the Heat, the overall probabilities of shooting the ball (sum of Make 2, Miss 2, Make 3, and Miss 3) are $9.1\%$ for cluster 1, $23\%$ for cluster 2, and $21.9\%$ for cluster 3. Meanwhile, the corresponding numbers for the Celtics are $14.0\%$ for cluster 1, $33.6\%$ for cluster 2, and $35.4\%$ for cluster 3. Relatively speaking, when releasing the ball, the Heat players have lower chances to take a shot than the Celtics players do, but higher chances to pass the ball to their teammates. This shows the offense of the Heat involves more interactions among players. For both teams, the respective shooting probabilities for clusters 2 and 3 are more than twice as high as those for cluster 1. Let us look into these probabilities in even greater detail. James and Wade (cluster 2, Heat) shoot many more 2-pointers than 3-pointers, and incredibly, they score more than half of their 2-pointer shots. Indeed, James and Wade are outstanding at penetration, but not great 3-point shooters. By contrast, Pierce and Allen (cluster 2, Celtics) are better balanced. They shoot and make more 3-pointers than James and Wade do. In the offensive end, Pierce has been regarded as one of the most well-rounded players (as of 2012), because of his ability to score from almost any location. Allen is an extraordinary 3-point shooter --- actually one of the best in the entire NBA history. Unfortunately, Pierce and Allen miss many 2-pointers in these two games. For the Heat, both their cluster 1 and cluster 3 shoot many 3-pointers (almost as many as 2-pointers), since one of their main strategies is for James and Wade to attract the defense from their opponents while their other players seek open-shot opportunities (mostly 3-pointers). For the Celtics, their clusters 1 and 3 mostly shoot 2-pointers, and their main attacking areas are close to the hoop.

Finally, we examine the probabilities of drawing a foul and committing a turnover.   Note that ``drawing a foul'' means being fouled by the opposing team, often after fooling them with fake moves.  For the Heat, James and Wade draw fouls with much higher probability than do their teammates in cluster 1 and cluster 3 ($6.3\%$ vs. $2.1\%$ and $1.3\%$). The reason is that James and Wade are often the ones to penetrate, while their teammates usually play ``catch and shoot''. For the Celtics, players in their cluster 3 have the highest probability of drawing fouls, because those players --- for example, Kevin Garnett --- are very aggressive when playing close to the hoop; players in their cluster 2 are also good at drawing fouls, as Pierce is a master at doing so. Overall, the Celtics are more capable of drawing fouls, but they make fewer turnovers than the Heat, because they play at a slower pace and make fewer passes.

\subsection{Cleveland Cavaliers versus Golden State Warriors in 2015} \label{sec.GC}
We now analyze two games in the 2015 NBA finals between the Cleveland Cavaliers and the Golden State Warriors --- in particular, games 2 and 5. Again, we consider only the first three quarters. These two games are particularly interesting case-study materials for us because there was a fascinating change in the Warriors' lineup in between. After losing both games 2 and 3 of the series, Steve Kerr, the head coach of the Warriors, decided to change their regular lineup to a small lineup, which meant that they stopped playing centers. This was an unconventional strategy but it successfully turned the series around, and the Warriors went on to win the championship that year by winning three consecutive games!

These two teams have very different styles of play to start with. The aforementioned change in the Warriors' lineup meant there was a big change in how the two teams played these two particular games as well. Thus, unlike in the previous section, in this section we simply fit four CSBMs separately for each team and each game, and no longer fit a pooled model combining the two teams and the two games together. Overall, there are four data sets. For game 2, the Cavaliers have eight players, 84 plays and 290 transactions, while the Warriors have ten players, 75 plays and 307 transactions. For game 5, the Warriors have ten players, 79 plays and 296 transactions, whereas the Cavaliers have eight players, 81 plays and 291 transactions. As in the previous section, in what follows we give detailed discussions about the clustering results, initial probability estimates, fitted rate functions, and transition probability estimates, in that order.

\paragraph{Clustering results} The cluster labels of the players for the two games are reported in Table \ref{cluster.g25}. As in the previous section, we set $K=3$ here as well.

\begin{table}[h]
\caption{Clustering results for the 2014-2015 Cleveland Cavaliers and Golden State Warriors ($K=3$). Cluster labels are C1, C2, C3.  Four different clustering results are presented (two teams $\times$ two games). Player positions are included for reference only; they are not used by the clustering algorithm.}  \label{cluster.g25}
\begin{center}
\begin{tabular}{|c|l|c|c|c|c|c|c|c|} \toprule
& & & \multicolumn{3}{ c| }{Game 2} & \multicolumn{3}{ c| }{Game 5} \\
\midrule
& & & \multicolumn{3}{ c| }{Alone} & \multicolumn{3}{ c| }{Alone} \\
\midrule
Team & \multicolumn{1}{ c| }{Player} & Position & C1 & C2 & C3 & C1 & C2 & C3 \\
\midrule
\multirow{8}{*}{\makecell{Cavaliers }}
& Matthew Dellavedova & PG &  & X  & & & X&\\
& Iman Shumpert & SG &  &X&  & &  &X\\
& J.R. Smith & SG &  & &X  & & X&\\
& LeBron James & SF &X & &   &X & & \\
& James Jones & SF & & &X  & & &X \\
& Mike Miller & SF & & &X & & &X\\
& Tristan Thompson & PF & & &X & & &X  \\
& Timofey Mozgov  & C & & &X & & &X\\ [0.3cm]
\cline{1-9}
  &&&&&&&&\\
\multirow{11}{*}{\makecell{Warriors }}
 & Stephen Curry & PG & X & &  &X & &\\
 & Shaun Livingston & PG & X & & &X & & \\
 & Klay Thompson & SG &  &X &   &  &X & \\
 & Leandro Barbosa & SG &  &X &  & &X  & \\
 & Harrison Barnes & SF &  & &X  & & X &\\
 & Andre Iguodala & SF & X & &   & & &X\\
 & Draymond Green & PF & X & &  & & &X\\
 & David Lee & PF &  \multicolumn{3}{ c| }{Did~~Not~~Play}  & &X &\\
 & Andrew Bogut & C &  & &X  & \multicolumn{3}{ c| }{Did~~Not~~Play}  \\
 & Festus Ezeli & C &  & &X   & \multicolumn{3}{ c| }{Did~~Not~~Play} \\
 & Marreese Speights & C &  & &X & \multicolumn{3}{ c| }{Did~~Not~~Play}  \\ [0.3cm]

\bottomrule
\end{tabular}
\end{center}
\end{table}

For the Cavaliers, the results from the two games are similar, except their two shooting guards --- Iman Shumpert and J.R. Smith --- switch clusters. It is not surprising that LeBron James is in a cluster by himself. In these two games, he is the only core player of the Cavaliers since their other two superstars, Kyrie Irving and Kevin Love, are both absent due to injuries. Without support from other superstar teammates, James has to take charge of a large amount of ball handling, passing and scoring; he simply does it all. Indeed, James is one of the most versatile players in the history of the NBA. With James being the only {\em primary} ball handler of the Cavaliers, their cluster 2 consists of {\em secondary} ball handlers: the point guard, Matthew Dellavedova, for both games; and a shooting guard --- Shumpert for game 2 and Smith for game 5. In general, both Shumpert and Smith can dribble and shoot. Shumpert handles the ball more often than does Smith in game 2, but their roles are reversed in game 5. Other than Smith (in game 2) and Shumpert (in game 5), their cluster 3 consists of \{James Jones, Mike Miller\}, both catch-and-shoot players, and \{Tristan Thompson, Timofey Mozgov\}, both inside (the paint) players. Overall, the Cavaliers are a team built around a single key player, LeBron James.

The Warriors, on the other hand, play the two games in fairly different styles. First of all, the active rosters are different: all three centers --- Andrew Bogut, Festus Ezeli and Marreese Speights --- play in game 2 but not in game 5; meanwhile, David Lee does not play in game 2, but does play in game 5.
We already explained the reason behind these changes in their lineup at the beginning of this section (Section~\ref{sec.GC}).
Beyond the clear change of rosters, our CSBM reveals more insight into the different playing styles of the Warriors in these two games. Unlike the Cavaliers, the Warriors have 4 primary ball handlers and distributors: Stephen Curry (PG), Shaun Livingston (PG), Andre Iguodala (SF) and Draymond Green (PF). In game 2 under their regular lineup, our model clusters these four players together. The two shooting guards, Klay Thompson and Leandro Barbosa, are clustered in one cluster. The three centers together with a small forward, Harrison Barnes, form the last cluster. In game 5 under their small lineup, our model divides their 4 primary ball handlers into two clusters --- the two point guards, Curry and Livingston, are in one cluster; the two forwards, Iguodala and Green, are in another. All remaining players are in a separate cluster. Note that, although both Barnes and Lee are forwards, their roles in the team are considerably less important than those of Iguodala and Green.

\paragraph{Initial probabilities}
\begin{table}[h]
\caption{Estimated transition probabilities ($P_{sk}$) from each initial action to clusters C1, C2 and C3, for four different clustering models of the 2014-2015 Cleveland Cavaliers and Golden State Warriors.} \label{ini.GC}
\begin{tabular}{c|l|ccc} \toprule
 & & C1 & C2 & C3\\
\midrule
\multirow{3}{*}{\makecell{Cavaliers \\ Game 2}} &Inbound & 0.489 & 0.422 & 0.089\\

& Rebound & 0.265 & 0.088 & 0.647\\

& Steal & 0.200 & 0.400 & 0.400\\
\midrule
\multirow{3}{*}{\makecell{Cavaliers \\ Game 5}} &Inbound & 0.500 & 0.409 & 0.091\\

& Rebound & 0.429 & 0.107 & 0.464\\

& Steal & 0.286 & 0.428 & 0.286\\
\midrule
\multirow{3}{*}{\makecell{Warriors \\ Game 2}} &Inbound & 0.767 & 0.093 & 0.140\\

& Rebound & 0.400 & 0.160 & 0.440\\

& Steal & 0.714 & 0.143 & 0.143\\
\midrule
\multirow{3}{*}{\makecell{Warriors \\ Game 5}} &Inbound & 0.660 & 0.140 & 0.200\\

& Rebound & 0.185 & 0.296 & 0.519\\

& Steal & 0.250 & 0 & 0.750\\
\bottomrule
\end{tabular}
\end{table}

The estimated transition probabilities from each initial action to the three clusters are shown in Table \ref{ini.GC}.

For the Cavaliers, the probabilities of the two games are similar, except the rebounds of LeBron James (the only player in cluster 1). James catches many more rebounds in game 5 than he does in game 2 (42.9\% vs. 26.5\%). The reason here is that, with the Warriors playing the small lineup, James becomes one of the tallest and biggest men on the court, playing closer to the rim and catching more rebounds. For both games, more than 90\% of the inbounds go to cluster 1 and cluster 2, with cluster 1 receiving slightly more than cluster 2. Players in cluster 2 contribute more than 40\% of the steals in the two games, while the other two clusters split the remainder.

For the Warriors, recall that their three centers, belonging to cluster 3 in game 2, do not play in game 5, and their two forwards, Iguodala and Green, belonging to cluster 1 in game 2, become the new cluster 3 in game 5. As a result, their inbound probabilities change slightly, but their rebound and steal probabilities change dramatically.
To get into more details,
their players in cluster 1 have a much higher probability of receiving an inbound than those in the other two clusters combined, because their cluster 1 contains two point guards, Curry and Livingston. However, this probability goes down by about 10\% from game 2 (76.7\%) to game 5 (66\%), whereas those of cluster 2 and cluster 3 each increases about 5\%. These results imply that, when the Warriors switch to their small lineup in game 5, players other than those in cluster 1 also get more opportunities to receive inbounds and initiate plays. In game 5, due to the absence of centers, who make up cluster 3 and contribute 44\% of the rebounds in game 2, all players start to share their contributions to catching rebounds as well. In particular, Green and Iguodala (in cluster 3 for game 5) now catch 51.9\% of the rebounds, in contrast to $<40\%$ when they are in cluster 1 for game 2; the contribution of cluster 2 to rebounds increases from 16\% in game 2 to 29.6\% in game 5; and finally, without Green and Iguodala (now in cluster 3), the two point guards that remain in cluster 1 (i.e., Curry and Livingston) also manage to catch 18.5\% of the rebounds. Regarding steals, the most significant changes are a huge decrease for cluster 1 (71.4\% to 25\%) and a huge boost for cluster 3 (14.3\% to 75\%). Once more, this is because Green and Iguodala have ``moved'' from cluster 1 to cluster 3; they both are top defenders who contribute to many steals.

\paragraph{Rate functions}
\begin{figure} [h]
\begin{center}
\includegraphics[width=\textwidth]{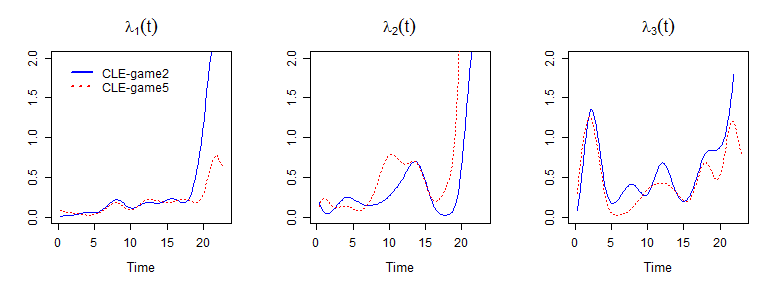}
\caption{Fitted rate functions for the 2014-2015 Cavaliers, $\lambda_1(t)$, $\lambda_2(t)$ and $\lambda_3(t)$, each describing the rates with which the ball leaves a player in cluster 1, cluster 2 and cluster 3, respectively.} \label{CLErate}
\end{center}
\end{figure}

\begin{figure} [h]
\begin{center}
\includegraphics[width=\textwidth]{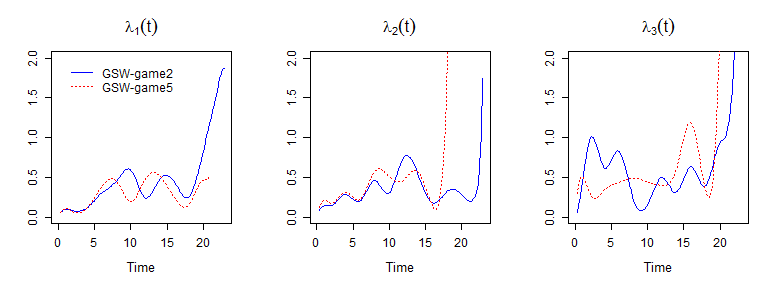}
\caption{Fitted rate functions for the 2014-2015 Golden State Warriors, $\lambda_1(t)$, $\lambda_2(t)$ and $\lambda_3(t)$, each describing the rates with which the ball leaves a player in cluster 1, cluster 2 and cluster 3, respectively.} \label{GSWrate}
\end{center}
\end{figure}

The fitted rate functions of the Cavaliers and the Warriors are displayed in Figure \ref{CLErate} and Figure \ref{GSWrate}, respectively.

For the Cavaliers, the rate functions from the two games appear to be generally similar for each respective cluster, with some small differences. For cluster 1 (James), its rate function $\lambda_1(t)$ is almost the same in the two games for $t<17$ --- fairly flat and low. This means that James plays with almost the same style at the beginning of a play in both games, keeping the ball in his hands and organizing the offense. Toward the end of a play, James starts to ``heat up'' at around $t\approx17$ in game 2, whereas he does so slightly later in game 5, at about $t\approx19$. This is because the small lineup of the Warriors in game 5 move much more quickly, so they can defend James more effectively in the last few seconds and delay his offense. For cluster 2, the first big difference appears after $t>7$. In game 2, $\lambda_2(t)$ grows {\em slowly} to reach a peak at $t\approx14$; however, in game 5, the same function $\lambda_2(t)$ grows {\em rapidly} after $t>7$ and maintains a high level until $t\approx14$. Clearly, players in cluster 2 have increased their offensive pace in game 5. On the one hand, Smith (cluster 2 SG in game 5) does more quick-release shooting than does Shumpert (cluster 2 SG in game 2). On the other hand, the higher defensive pressure created by the Warriors' small lineup has forced the Cavaliers to move the ball more quickly. For the same reasons, toward the end of a play, players in cluster 2 also tend to attack the rim or pass the ball slightly earlier in game 5 (at $t\approx16$) than they do in game 2 (at $t\approx18$). For cluster 3, their rate function $\lambda_3(t)$ displays a similar pattern in the two games, but the one in game 5 is almost entirely dominated by the one in game 2. Players in this cluster are big men and typically catch-and-shoot players; they are usually not responsible for handling the ball. They catch rebounds and start a play by passing the ball to their teammates in the other two clusters. At around $t\approx12$, they get their first chance to touch the ball, when they either shoot or pass it back to the ball handlers. Their second chance to touch the ball happens near the end of a play, when they have to shoot rapidly. In game 5, the small lineup of the Warriors can quickly cover the open shots and ``double up'' to defend a big man in the paint, and that forces players in Cavaliers' cluster 3 to keep the ball in their hands for a slightly longer period. This is why their $\lambda_3(t)$ is lower in game 5 than in game 2. Overall, the patterns displayed in the Cavaliers' three rate functions are quite similar in the two games. The changes mostly can be attributed to the different defensive strategies used by their opponent.

For the Warriors, though, due to the change in their lineup, their rate functions from the two games are noticeably different. In game 2 with their regular lineup, their rate functions (blue solid lines in Figure~\ref{GSWrate}) show regular patterns --- at the start of a play, $\lambda_1(t)$ and $\lambda_2(t)$ are relatively low, while $\lambda_3(t)$ has high peaks. This means that, at the start of a play, players in cluster 1 and cluster 2 tend to handle the ball, whereas those in cluster 3 catch rebounds and pass the ball out more or less immediately. This is the same as the playing style of the Cavaliers. However, their rate functions have more peaks than those of the Cavaliers. Moreover, their $\lambda_1(t)$ and $\lambda_2(t)$ in game 2 are, in general, higher than those of the Cavaliers at the start of a play. These show that the Warriors' offense is more flexible --- the ball is passed more frequently, so everybody gets chances to touch it, and no one holds the ball for a very long time. In fact, this has become the Warriors' signature team-playing style. However, in game 5, all three of their rate functions show significant differences. First, the two peaks of $\lambda_1(t)$ occur earlier in game 5 than in game 2. Second, the rapid growth of $\lambda_2(t)$ also appears earlier in game 5 (at $t\approx17$) than in game 2 (at $t\approx22$). Both differences indicate that, with a small lineup, the Warriors have increased their offensive pace in game 5. Finally, their $\lambda_3(t)$ changes dramatically between the two games; in game 5, it is much flatter at the beginning and has a much higher peak at $t\approx17$. This is certainly because, in game 5, the players making up cluster 3 are entirely different from the ones in game 2. From Table \ref{ini.GC}, we know that their cluster 3 in game 5 (Green and Iguodala) catch a larger proportion of rebounds than do their cluster 3 in game 2 (three centers), but instead of immediately passing the ball out, Green and Iguodala both often dribble and run the play. The peak of $\lambda_3(t)$ at $t\approx17$ in game 5 is particularly significant, revealing one key offensive strategy of the Warriors' small lineup, the so-called ``high pick-and-roll''. A typical sequence of this strategy is as follows: Curry dribbles the ball outside the three-point line, and Green (or Iguodala) comes to set up a screen (a ``human body wall''). Thanks to Curry's incredible three-point shooting skills, after he dribbles around the screen both defenders of Curry and of Green (or Iguodala) usually have to focus on covering Curry together, leaving Green (or Iguodala) wide open, so Curry can now pass the ball to him. Green (or Iguodala) can then shoot the ball; drive to the basket directly; or take one or two dribbles, draw another defender, and then pass the ball to another wide-open teammate, who is usually waiting at the three-point line on the weakly-defended side. This entire sequence often happens very quickly within three seconds.

Overall, the estimated rate functions reveal many intricate details of a team's playing style. The Cavaliers play around their key superstar, LeBron James, whereas the Warriors share the ball more evenly. It also can be easily seen that the Warriors have played these two games quite differently and the Cavaliers have responded with small but clear adjustments in their playing style as well.

\paragraph{Transition probabilities}

\begin{table}[h]
\caption{Estimated transition probabilities ($P_{kl}$ and $P_{ka}$) for the 2014-2015 Cleveland Cavaliers and Golden State Warriors ($K=3$).  Rows are originating clusters and columns are receiving clusters and play outcomes.} \label{tran.GC}
\begin{center}
\scriptsize
\begin{tabular}{c|c|ccccccccc} \toprule
 && C1 & C2 & C3 & Make2 & Miss2 & Make3 & Miss3 & Fouled & TO\\
\midrule
\multirow{3}{*}{\makecell{Cavaliers \\ Game 2}} & C1 & 0 & 0.167 & 0.430  & 0.111 & 0.153 & 0.014 & 0.014 & 0.069 & 0.042\\

 & C2 & 0.292 & 0.141 & 0.259 & 0.016  & 0.081 & 0 & 0.114 & 0.032 & 0.065\\

& C3 & 0.335 & 0.160 & 0.111 & 0.098  & 0.123 & 0.037 & 0.037 & 0.062 & 0.037\\
\midrule
\multirow{3}{*}{\makecell{Cavaliers \\ Game 5}} & C1 & 0 & 0.384 & 0.274 & 0.123  & 0.110 & 0 & 0.027 & 0.027 & 0.055\\

 & C2 & 0.296 & 0.181 & 0.261 & 0.024  & 0.036 & 0.059 & 0.107 & 0 & 0.036\\

& C3 & 0.346 & 0.198 & 0.076 & 0.061  & 0.122 & 0.045 & 0.061 & 0.061 & 0.030\\
\midrule
\multirow{3}{*}{\makecell{Warriors \\ Game 2}} & C1 & 0.357 & 0.233 & 0.194 & 0.037  & 0.037 & 0.007 & 0.060 & 0.030 & 0.045\\

 & C2 & 0.380 & 0 & 0.120 & 0.140  & 0.080 & 0.100 & 0.120 & 0.060 & 0\\

& C3 & 0.469 & 0.226 & 0 & 0.061  & 0.081 & 0 & 0.061 & 0.061 & 0.041\\
\midrule
\multirow{3}{*}{\makecell{Warriors \\ Game 5}} & C1 & 0.159 & 0.276 & 0.323 & 0.058  & 0.058 & 0.046 & 0.034 & 0 & 0.046\\

 & C2 & 0.151 & 0.097 & 0.285 & 0.151  & 0.166 & 0.015 & 0.015 & 0.060 & 0.060\\

& C3 & 0.330 & 0.267 & 0.137 & 0.064  & 0.038 & 0.025 & 0.038 & 0.076 & 0.025\\
\bottomrule
\end{tabular}
\end{center}
\end{table}

The estimated transition probabilities of events originating from the three different clusters are displayed in Table \ref{tran.GC}, for both the Cavaliers and the Warriors.

For the Cavaliers, the overall probabilities to pass the ball (sum of the first three columns) for the three respective clusters are \{59.7\%, 69.2\%, 60.6\%\} in game 2, and \{65.8\%, 73.8\%, 62\%\} in game 5. Clearly, the Cavaliers make more passes in game 5 than in game 2, which is due to the stronger defense by the Warriors' small lineup. For the same reason, in game 2 James (the only player in cluster 1) passes more to cluster 3 (shooters and big men), whereas in game 5 he passes more to cluster 2 (ball handlers). The respective roles of their cluster 2 and cluster 3 do not change much in the two games --- cluster 2 is the bridge between cluster 1 and cluster 3, making almost an equal proportion of passes to each of the other two clusters; cluster 3, however, more often passes the ball back to James (cluster 1). The overall probabilities to shoot the ball (sum of columns 4-7) for the three respective clusters are \{29.2\%, 21.1\%, 29.5\%\} in game 2, and \{26\%, 22.6\%, 28.9\%\} in game 5, which do not change much. When facing the quick defense of the Warriors in game 5, the Cavaliers has successfully created an almost equal percentage of shots by making more passes. Regarding the probabilities of being fouled and making turnovers, James (cluster 1) fails to draw as many fouls in game 5 as he does in game 2 (2.7\% vs. 6.9\%), but he makes more turnovers (5.5\% vs. 4.2\%). These can be partly attributed, again, to the stronger defense by the Warriors' small lineup, especially the one-on-one defense on James by Iguodala. Players in cluster 2 are not as aggressive in game 5 as they are in game 2 --- although they make fewer turnovers (3.6\% vs. 6.5\%), they do not draw any fouls at all (0\% vs. 3.2\%). The performance of cluster 3 is fairly stable in the two games in terms of drawing fouls and making turnovers.

For the Warriors, the overall passing probabilities of their three respective clusters are \{78.4\%, 50\%, 79.5\%\} in game 2, and \{75.8\%, 53.3\%, 73.4\%\} in game 5. Despite the drastic changes in their lineup, these probabilities do not change much. 
Each of the first three columns in Table~\ref{tran.GC} contains the probabilities that the corresponding cluster is the receiver of the ball passed from different clusters. Here, we can easily see that a considerable proportion of the passes have shifted from cluster 1 to cluster 3 in game 5. This is because Green and Iguodala, two of the four primary ball handlers, are now in cluster 3 as opposed to cluster 1, and they receive many passes. The overall shooting probabilities (sum of columns 4-7) for their three respective clusters are \{14.1\%, 44\%, 20.3\%\} in game 2, and \{19.6\%, 34.7\%, 13.5\%\} in game 5. In both games, players in cluster 2 are more likely to shoot than those in the other two clusters. This makes sense because cluster 2 contains two shooting guards, Klay Thompson and Leandro Barbosa, who both are excellent scorers and often take on a huge responsibility in shooting the ball. It can also be seen that, in game 5, the probability to shoot has increased for cluster 1 but decreased for cluster 2. This is because the small lineup gives players in cluster 1 --- especially Curry --- more open space and hence better shooting opportunities; by contrast, Klay Thompson (cluster 2), who is less affected by the change in the lineup, struggles with shooting in game 5. For cluster 3, we see that Green and Iguodala (cluster 3 in game 5) are less likely to shoot than the centers (cluster 3 in game 2). With regard to shooting, it is well-known that the Warriors rely on three-pointers as one of their most important scoring methods. Curry and Thompson are arguably the best three-point shooting back-court duo in the entire history of the NBA. From Table \ref{tran.GC}, we can clearly see that the Warriors attempt many more three-pointers than the Cavaliers do and they also succeed more often. One surprising observation is that players in their cluster 2 shoot considerably fewer three-pointers in game 5 than they do in game 2. Indeed, this is another piece of evidence showing the struggle of Klay Thompson in game 5. There are two significant differences in terms of drawing fouls and making turnovers: cluster 1 fails to draw any fouls in game 5 versus 3\% in game 2; and cluster 2 makes more turnovers in game 5 than in game 2 (6\% vs. 0\%).

\subsection{LeBron James: Miami Heat versus Cleveland Cavaliers} \label{sec.HCa}
Both the 2011-12 Miami Heat and the 2014-15 Cleveland Cavaliers had LeBron James on their teams and made him the key player. Thus, it is especially interesting for us to compare the player structures of these two teams, and to see if there is any difference in how James has played the game with these different teams. We investigate the first question by pooling the transactions of the Heat (in their two 2012 games versus the Celtics) and the transactions of the Cavaliers (in their two 2015 games versus the Warriors), and applying the CSBM to cluster the players from both teams together. With regard to the second question, we simply compare the individual results we have obtained earlier for the Heat (Section \ref{sec.HC}) and for the Cavaliers (Section \ref{sec.GC}).

For the pooled CSBM, we focus primarily on the clustering results in this section and forsake any detailed discussions of the rate functions or the transition probabilities. Other than LeBron James, Mike Miller and James Jones are also on both of these teams. When playing on different teams, the same player may play in a different style, depending on his specific role for the team. Hence for James (and likewise for Miller and Jones, too), we create two separate avatars --- one for the games he played on the Heat and another for the games he played on the Cavaliers --- and treat them as two different ``players'' in the clustering algorithm. We are especially curious whether the pooled CSBM will cluster the two avatars of the same player into the same cluster or different clusters.

Table \ref{cluster.HC} displays the clustering results from the pooled CSBM, fitted to all transactions of the Heat and the Cavaliers in the 4 games we have annotated.  With a total of 19 ``players'', we now choose $K=4$ instead of $K=3$ as we did in the previous two sections; this allows us to cluster the ``players'' with a slightly finer resolution.

Our clustering results clearly indicate that the 2011-12 Heat and the 2014-15 Cavaliers are built in a very similar way. Cluster 1 consists of point guards; cluster 2 consists of superstars --- namely, LeBron James (for both teams) and Dwyane Wade (for the Heat); cluster 3 consists of the other perimeter players --- mostly shooters and perimeter defenders; and the last cluster is made up of big men --- power forwards and centers. It also turns out that the two avatars of the same player (whether James, Miller or Jones) are always clustered together. Indeed, both teams are built around LeBron James and their playing styles are similar, too. James is the primary ball handler and distributor for both teams. While playing for the Heat, James has Wade as an important helper, but while playing for the Cavaliers, he is the only superstar. We can imagine that, if Kyrie Irving, the superstar point guard of the Cavaliers, were not injured, he might have joined James and Wade in  cluster 2. The point guards in these two team are secondary ball handlers and serve as bridges between the superstars and the other players. Players in cluster 3 are mainly responsible for playing defense and ``catch and shoot''. The big men in cluster 4 are mostly responsible for catching rebounds and scoring under the rim.

\newcolumntype{$}{>{\global\let\currentrowstyle\relax}}
\newcolumntype{^}{>{\currentrowstyle}}
\newcommand{\rowstyle}[1]{\gdef\currentrowstyle{#1}%
  #1\ignorespaces
}
\begin{table}[h]
\caption{Clustering results for the 2011-2012 Miami Heat and the 2014-2015 Cleveland Cavaliers together ($K=4$).  Cluster labels are C1, C2, C3, C4. Players appearing with two separate avatars for the clustering algorithm are bolded. Player positions are included for reference only; they are not used by the clustering algorithm.} \label{cluster.HC}
\begin{center}
\begin{tabular}{|$c|^l|^c|^c|^c|^c|^c|} \toprule
& & & \multicolumn{4}{ c| }{Together}  \\
\midrule
Team & \multicolumn{1}{ c| }{Player} & Position & C1 & C2 & C3 & C4 \\
\midrule
\multirow{11}{*}{Heat} & Mario Chalmers & PG & X & & &   \\
 & Norris Cole & PG & X & & &   \\
 & Dwyane Wade & SG &  &X & &   \\
\rowstyle{\bf} & LeBron James & SF &  &X & &   \\
\rowstyle{\bf} & James Jones & SG &  & &X &   \\
 & Shane Battier & SF &  & &X &   \\
\rowstyle{\bf} & Mike Miller & SF &  & &X &   \\
 & Chris Bosh & PF &  & & & X  \\
 & Udonis Haslem & PF &  & & & X  \\
 & Ronny Turiaf & C &  & & & X  \\
 & Joel Anthony & C &  & &  & X  \\
  &&&&&&\\
\multirow{8}{*}{\makecell{Cavaliers }}
& Matthew Dellavedova & PG & X &   & & \\
& Iman Shumpert & SG &  & & X & \\
& J.R. Smith & SG &  & &X  & \\
\rowstyle{\bf} & LeBron James & SF & &X &   &  \\
\rowstyle{\bf} & James Jones & SF & & &X  &  \\
\rowstyle{\bf} & Mike Miller & SF & & &X & \\
& Tristan Thompson & PF & & & & X  \\
& Timofey Mozgov  & C & & & & X\\ [0.3cm]
\bottomrule
\end{tabular}
\end{center}
\end{table}

In the rest of this section, we revisit some individual results for the Heat (Section \ref{sec.HC}) as well as for the Cavaliers (Section \ref{sec.GC}) in order to compare in more detail the performance of LeBron James in those two series.

First, recall that our cluster labels (e.g., C1, C2, ...) are arbitrary, and that James has been clustered into C2 with the 2011-12 Heat but into C1 with the 2014-15 Cavaliers. Comparing Figure \ref{Rate.k3} (middle panel) and Figure \ref{CLErate} (left panel), we find that the Heat's $\lambda_2(t)$ function has more peaks and is higher than the Cavaliers' $\lambda_1(t)$ function overall. This shows that, while playing for the Heat, James chooses to pass the ball more often at the beginning of a play. This is mostly because of the presence of Wade, a superstar teammate, who interacts with James more frequently than the point guards. Actually, the ``two-man fast break'' by James and Wade is one of the Heat's defining features. Second, comparing Table \ref{ini.k3} and Table \ref{ini.GC}, we find that, while playing for the Heat, James and Wade {\em together} receive 19.4\% of the inbounds, whereas, while playing for the Cavaliers, James {\em alone} receives a staggering 50\% of the inbounds. The Heat mostly let their point guards carry the ball past the half court, because they always have one of them (either Chalmers or Cole) on the court. However, with Irving out on injury, the Cavaliers only play one point guard (Dellavedova) in their lineup, so James has to carry the ball more than usual. Third, while on the Heat, James and Wade {\em together} average a 23\% probability to shoot, but while on the Cavaliers, James {\em alone} has an even higher probability to shoot --- 29.2\% and 26\% respectively in the two games against the Warriors. James is a great scorer as well as offensive organizer. He can freely switch between these two modes of play depending on the situations in the game. While playing for the 2011-12 Heat, James has stronger teammates, so he tends to create more shooting opportunities for others. With the 2014-15 Cavaliers, however, James must take more shots by himself due to the limited support from his teammates.

In summary, our analysis using the CSBM shows that the player structure of the 2011-12 Heat and that of the 2014-15 Cavaliers are fairly similar. The CSBM also reveals many subtle differences in LeBron James' playing style in the two series.

%% file: Section6-MZ.tex
\section{Summary}

In this paper, we advocate the concept that basketball games can be analyzed as transactional networks. We have proposed a Continuous-time Stochastic Block Model to cluster players based on their styles of handling the ball. In particular, we model each basketball play as an inhomogeneous continuous-time Markov chain, with transition rate functions being governed by the players' cluster membership. We adopt B-splines to model the rate functions and an EM$^+$ algorithm to estimate model parameters. Applications to a number of NBA games between the 2011-12 Miami Heat and Boston Celtics and between the 2014-15 Cleveland Cavaliers and Golden State Warriors have yielded compelling evidence that the CSBM framework is of great practical value in clustering and evaluating basketball players.


As the popularity of basketball analytics appears to be growing in recent years, it is perhaps helpful for us to summarize the main differences between our work and a few recent works in this area \citep[e.g.,][]{Bbnet,Bornn2}. The key features of our work are: (i) viewing basketball games from a network perspective, (ii) consideration of time dynamics, and (iii) clustering of players at an individual level. In what follows, we discuss how our work differs from a few others in terms of these features; a brief summary is given in Table~\ref{comp.others}.

\begin{table}[t]
\caption{Summary of differences between our work and others.} \label{comp.others}
\centering
\begin{tabular}{l|ccc} \toprule
        & Network     & Time     & Model \\
        & Perspective & Dynamics & Objective \\
\midrule
CSBM           & yes & yes &  descriptive at individual level\\
\citet{Bbnet}  & yes & no  &  descriptive at position level\\
\citet{Bornn2} & no  & yes &  predictive at individual level\\
               & & & (of final point outcome) \\
\bottomrule
\end{tabular}
\end{table}

\citet{Bbnet} certainly view basketball games from a network perspective as well, but they do not take time dynamics into account, and their treatment of players occurs at a position level rather than an individual level. Specifically, they pre-group players according to their on-court positions (e.g., point guard, shooting guard, and so on). Whereas our CSBM describes player differences based on the real-time dynamics of how each basketball play unfolds, the method developed by \citet{Bbnet} aims to describe differences in how each of the five pre-defined positions communicates with each other --- and with various initial and absorbing states --- at an aggregate level, aggregated over both all players holding the same position and all transactions during a certain time period (e.g., an entire game). In their work, point guards are always considered together with other point guards, and any player difference at the individual level is suppressed. While it is hardly surprising that many players holding the same position often end up being clustered together by our CSBM, this is certainly not always the case. For example, our analysis of the two games between the 2014-15 Cleveland Cavaliers and Golden State Warriors (Section~\ref{sec.GC}) clearly shows that the distinctive playing style of LeBron James almost calls for the definition/creation of a new on-court position, for which some long-time basketball observers have informally suggested the name of ``point forward''. Our analysis also shows that players like Draymond Green (a power forward) and Andre Iguodala (a small forward) are certainly playing important roles in the game beyond the traditional ones defined by their respective on-court positions.

\citet{Bornn2}, on the other hand, do consider time dynamics, but they do not view basketball games from a network perspective. While they track the movement of the ball both spatially and over time, they do not view players as nodes and passes as edges. Most importantly, their objective is fundamentally different from ours. Our goal is to cluster players according to their individual playing styles as characterized by the rate functions $\lambda_k(t)$ and the transition probabilities $P_{sk}$, $P_{kl}$, and $P_{ka}$, but theirs is to predict the final point value of each basketball play/possession as the individual play unfolds. One can say that their analysis is driven by {\em outcome} but ours is driven by {\em style}.
Although rate functions for ball passing are components of both models, their structure and role vary considerably. Our rate functions are smooth functions of clock time, and are used to characterize groups of players with similar transition rates. Their rate functions are log-linear regressions which use predictors derived from motion-capture data, forming one component in a hierarchical model whose ultimate objective is to predict point value. They are not used to cluster players.

Early works on the SBM \citep{Snijders1997,firststep} are mostly concerned with static networks. Recently, \citet{Ho2011}, and \citet{XuHero2014} have used dynamic SBMs to study social networks that evolve over time, but their works focus on {\em discrete} time dynamics and are thus not directly applicable to network transactions (such as basketball passes) that happen in {\em continuous} time. Although we have focused on basketball games in this paper, one certainly can use our CSBM to analyze any other network where exchanges take place between its nodes in continuous time.


%% file: Appendix-MZb-rev.tex

\section{Some details about inhomogeneous Poisson processes}
\label{appdx:Poisson}

Consider an inhomogenous Poisson process with rate function $\rho(t)$. We derive the distribution of having $m$ events arriving at time points $t_1<t_2<\ldots<t_m \in [T_0, T]$, closely following the presentation by \citet[p.~30]{Cook.book}.
Let $N_t$ denote the number of events in the time interval $[t,t+\Delta t)$. By the definition of the Poisson process, for a very small $\Delta t$,
\begin{align}
\mathcal{P}(N_t=0)&=1-\rho(t)\Delta t+o(\Delta t), \\
\mathcal{P}(N_t=1)&=\rho(t)\Delta t +o(\Delta t),\\
\mathcal{P}(N_t\geq 2)&=o(\Delta t).
\end{align}
Consider a partition of $[T_0,T)$, say $T_0=u_0 < u_1 < u_2\cdots<u_R=T$. By the ``independent increment'' property of the Poisson process, we have
\begin{multline}
\mathcal{P}([T_0, T_1))=\prod_{r=0}^{R-1}\mathcal{P}([u_r,u_{r+1}))=\prod_{r=0}^{R-1}\mathcal{P}(N_{u_r}) \\
= \Big(\prod_{N_{u_r}=0} [1-\rho(u_r)\Delta u_r + o(\Delta u_r)] \Big)  \cdot
  \Big(\prod_{N_{u_r}=1} [\rho(u_r)\Delta u_r + o(\Delta u_r)] \Big) \cdot \\
  \Big(\prod_{N_{u_r}\geq 2} [o(\Delta u_r)] \Big) \label{last2}.
\end{multline}
Notice that $\log[1-\rho(t)\Delta t]=-\rho(t)\Delta t+o(\Delta t)$, so the logarithm of the first product in \eqref{last2} --- the one over $N_{u_r}=0$ --- approaches the Riemann integral, $-\int_{T_0}^T \rho(t)dt$, in the limit.
Thus, dividing $\Delta u_r$ into each respective term that corresponds to the interval $[u_r,u_{r+1})$ and taking the limit as $R\rightarrow \infty$ and consequently as $\Delta u_r = u_{r+1} - u_r \rightarrow 0$, we obtain that the desired distribution is
\[
\prod_{i=1}^m \rho(t_i)\cdot \exp\big[{-\int_{T_0}^{T}\rho(u)du}\big].
\]

\section{Some details for the EM algorithm}

\subsection{The conditional expectation $\mathbf{E}[\log\mathcal{L}(\mathbf{T,Z})|\mathbf{T};\Theta^*]$}
\label{appdx:condEdetail}

First, by \eqref{conlike.bnet}, we have
\begin{align}
&\mathbf{E}[\log\mathcal{L}(\mathbf{T,Z})|\mathbf{T};\Theta^*]  \label{estep.t} \\
&=\mathbf{E}[\log\mathcal{L}(\mathbf{T|Z})+\log\mathcal{L}(\mathbf{Z})|\mathbf{T};\Theta^*] \nonumber \\
&=\mathbf{E}\Big[
 \sum_{s\in\mathcal{S}}\sum_{i=1}^n\log\mathcal{L}^I(\mathbf{T}_{si}|\mathbf{Z})+
 \sum_{1\leq i\neq j\leq n}\log\mathcal{L}^{P_1}(\mathbf{T}_{ij}|\mathbf{Z})   \nonumber \\
 & \quad \quad +\sum_{i=1}^n \log\mathcal{L}^{P_2}(\mathbf{T}_i|\mathbf{Z}) +\sum_{i=1}^{n}\sum_{a\in\mathcal{A}}\log\mathcal{L}^O(\mathbf{T}_{ia}|\mathbf{Z})+
 \log\mathcal{L}(\mathbf{Z})\Big|\mathbf{T};\Theta^*\Big] \nonumber
\end{align}
Now, we plug in \eqref{compo1}, \eqref{compo2}, \eqref{compo3}, \eqref{compo4} and \eqref{compo5}, and the respective terms in \eqref{estep.t} are as follows. The $\mathcal{L}^I$ part is equal to
\begin{align}
\sum_{s\in\mathcal{S}}\sum_{i=1}^n&\mathbf{E}\Big[\log\mathcal{L}^I(\mathbf{T}_{si}|\mathbf{Z})\Big|\mathbf{T};\Theta^*\Big] \label{estep1} \\
&=\sum_{s\in\mathcal{S}}\sum_{i=1}^n\mathbf{E}\Big[\log\prod_{k=1}^K\Big(\prod_{h=1}^{m_{si}}\big(P_{sk}\cdot \frac{1}{G_{k}^{sih}}\big)\Big)^{z_{ik}}\Big|\mathbf{T};\Theta^*\Big]  \nonumber \\
&=\sum_{s\in\mathcal{S}}\sum_{i=1}^n\sum_{k=1}^K\mathbf{E}\Big[z_{ik}\cdot \sum_{h=1}^{m_{si}}\big(\log P_{sk}- \log G_k^{sih}\big)\Big|\mathbf{T};\Theta^*\Big] \nonumber \\
&=\sum_{s\in\mathcal{S}}\sum_{i=1}^n\sum_{k=1}^K\Big(\mathbf{E}[z_{ik}|\mathbf{T};\Theta^*]\cdot m_{si}\log P_{sk}\Big)  \nonumber\\
& \qquad \qquad \qquad -\sum_{s\in\mathcal{S}}\sum_{i=1}^n\sum_{k=1}^K\mathbf{E}\Big[z_{ik}\cdot \sum_{h=1}^{m_{si}}\log G_k^{sih}\Big|\mathbf{T};\Theta^*\Big] \nonumber.
\end{align}
The $\mathcal{L}^{P_1}$ part is equal to
\begin{align}
&\sum_{1\leq i\neq j\leq n}\mathbf{E}\Big[\log\mathcal{L}^{P_1}(\mathbf{T}_{ij}|\mathbf{Z})\Big|\mathbf{T};\Theta^*\Big] \label{estep2} \\
&=\sum_{1\leq i\neq j\leq n}\mathbf{E}\Big[\log\prod_{k=1}^K\prod_{l=1}^K\bigg(\prod_{h=1}^{m_{ij}}\Big(\rho_{kl}(t_{ijh})\cdot \frac{1}{G_l^{ijh}}\Big)\bigg)^{z_{ik}z_{jl}}\Big|\mathbf{T};\Theta^*\Big] \nonumber \\
&=\sum_{1\leq i\neq j\leq n}\mathbf{E}\Big[\sum_{k=1}^K\sum_{l=1}^K\Big(z_{ik}z_{jl}\cdot \sum_{h=1}^{m_{ij}}\big(\log\rho_{kl}(t_{ijh}) -\log{G_l^{ijh}}\big)\Big)\Big|\mathbf{T};\Theta^*\Big] \nonumber \\
&=\sum_{1\leq i\neq j\leq n}\sum_{k=1}^K\sum_{l=1}^K\Big(\mathbf{E}[z_{ik}z_{jl}|\mathbf{T};\Theta^*]\cdot \sum_{h=1}^{m_{ij}}\log\rho_{kl}(t_{ijh})\Big) \nonumber \\
& \qquad \qquad \qquad \qquad - \sum_{1\leq i\neq j\leq n}\sum_{k=1}^K\sum_{l=1}^K\mathbf{E}\Big[z_{ik}z_{jl}\cdot \sum_{h=1}^{m_{ij}}\log{G_l^{ijh}}\Big|\mathbf{T};\Theta^*\Big]. \nonumber
\end{align}
The $\mathcal{L}^{P_2}$ part is equal to
\begin{align}
&\sum_{i=1}^n \mathbf{E}\Big[\log\mathcal{L}^{P_2}(\mathbf{T}_i|\mathbf{Z})\Big|\mathbf{T};\Theta^* \Big] \label{estep3} \\
&=\sum_{i=1}^n \mathbf{E}\Big[\log\prod_{k=1}^K\bigg(\prod_{h=1}^{M_i}\exp\Big(-\sum_{l=1}^K\int_{t_{ih}^-}^{t_{ih}}\rho_{kl}(t)\cdot I(G_l^{ih}>0)\mathrm{d}t\Big)\bigg)^{z_{ik}}\Big|\mathbf{T};\Theta^* \Big] \nonumber \\
&=\sum_{i=1}^n \sum_{k=1}^K\mathbf{E}\Big[z_{ik}\sum_{h=1}^{M_i}\Big(-\sum_{l=1}^K\int_{t_{ih}^-}^{t_{ih}}\rho_{kl}(t)\cdot I(G_l^{ih}>0)\mathrm{d}t\Big)\Big|\mathbf{T};\Theta^* \Big] \nonumber \\
&=-\sum_{i=1}^n \sum_{k=1}^K\sum_{h=1}^{M_i}\sum_{l=1}^K\mathbf{E}\Big[z_{ik}\int_{t_{ih}^-}^{t_{ih}}\rho_{kl}(t)\cdot I(G_l^{ih}>0)\mathrm{d}t \Big|\mathbf{T};\Theta^* \Big] \nonumber \\
&=-\sum_{i=1}^n \sum_{k=1}^K\sum_{l=1}^K\sum_{h=1}^{M_i}\Big(\mathbf{E}\Big[z_{ik}I(G_l^{ih}>0)\Big|\mathbf{T};\Theta^* \Big]\cdot \int_{t_{ih}^-}^{t_{ih}}\rho_{kl}(t) \mathrm{d}t\Big), \nonumber
\end{align}
where we have pulled the indicator term $I(G_{l}^{ih}>0)$ out of the integral in the last step of \eqref{estep3} because the quantity $G_l^{ih}$ is a constant on any $(t_{ih}^-,t_{ih}]$, as no player substitution can happen during that time. Finally, the $\mathcal{L}^O$ part is equal to
\begin{align}
&\sum_{i=1}^{n}\sum_{a\in\mathcal{A}}\mathbf{E}\Big[\log\mathcal{L}^O(\mathbf{T}_{ia}|\mathbf{Z})\Big|\mathbf{T};\Theta^* \Big] \label{estep4} \\
&=\sum_{i=1}^{n}\sum_{a\in\mathcal{A}}\mathbf{E}\Big[\log\prod_{k=1}^K\bigg(\prod_{h=1}^{m_{ia}}\eta_{ka}(t_{iah})\cdot \prod_{h=1}^{M_i}\exp\Big(-\int_{t_{ih}^-}^{t_{ih}}\eta_{ka}(t)\mathrm{d}t\Big)\bigg)^{z_{ik}}\Big|\mathbf{T};\Theta^* \Big] \nonumber \\
&=\sum_{i=1}^{n}\sum_{a\in\mathcal{A}}\sum_{k=1}^K\mathbf{E}\Big[z_{ik}\Big(\sum_{h=1}^{m_{ia}}\log\eta_{ka}(t_{iah})- \sum_{h=1}^{M_i}\int_{t_{ih}^-}^{t_{ih}}\eta_{ka}(t)\mathrm{d}t\Big)\Big|\mathbf{T};\Theta^* \Big] \nonumber \\
&=\sum_{i=1}^{n}\sum_{a\in\mathcal{A}}\sum_{k=1}^K\bigg(\mathbf{E}[z_{ik}|\mathbf{T};\Theta^* ]\cdot\Big(\sum_{h=1}^{m_{ia}}\log\eta_{ka}(t_{iah})- \sum_{h=1}^{M_i}\int_{t_{ih}^-}^{t_{ih}}\eta_{ka}(t)\mathrm{d}t\Big)\bigg), \nonumber
\end{align}
and the $\mathcal{L}(Z)$ part is equal to
\begin{align}
\mathbf{E}\Big[\log\mathcal{L}(\mathbf{Z})|\mathbf{T};\Theta^*\Big]&=
\mathbf{E}\Big[\log\prod_{i=1}^n\prod_{k=1}^K\pi_{k}^{z_{ik}}|\mathbf{T};\Theta^*\Big] \label{estep5} \\
&=\sum_{i=1}^n\sum_{k=1}^K\Big(\mathbf{E}[z_{ik}|\mathbf{T};\Theta^*]\cdot \log\pi_k\Big).\nonumber
\end{align}

\subsection{Analytic updates of $\bm{\pi}$ and $\mathbf{P}$}
\label{appdx:prob-updates}

In the conditional expectation of the log-likelihood \eqref{estep.t}, the term that contains $\mathbf{P}$, the transition probabilities from initial actions to players in different groups, appears in \eqref{estep1}. It is
\begin{equation}
\sum_{s\in\mathcal{S}}\sum_{i=1}^n\sum_{k=1}^K\Big(\mathbf{E}[z_{ik}|\mathbf{T};\Theta^*]\cdot m_{si}\log P_{sk}\Big)
\end{equation}
but there is a constraint
\begin{equation}\label{LGconstraintP}
\sum_{k=1}^KP_{sk}=1 \mbox{ for any } s\in\mathcal{S}.
\end{equation}
Introducing Lagrange multipliers $\zeta_s$, for each $s\in\mathcal{S}$, we get
\begin{equation}
\sum_{s\in\mathcal{S}}\bigg[
\sum_{i=1}^n\sum_{k=1}^K\Big(\mathbf{E}[z_{ik}|\mathbf{T};\Theta^*]\cdot m_{si}\log P_{sk}\Big)-\zeta_s\big(\sum_{k=1}^KP_{sk}-1 \big)\bigg].
\end{equation}
Differentiating with respect to each $P_{sk}$ and setting the the derivatives to zero, we get
\begin{equation}
\frac{\sum_{i=1}^n\big(\mathbf{E}[z_{ik}|\mathbf{T};\Theta^*]\cdot m_{si}\big)}{P_{sk}}-\zeta_s=0,  \mbox{ for } s\in\mathcal{S} \mbox{ and } k=1,2,\ldots,K.
\end{equation}
The constraint \eqref{LGconstraintP} implies
\begin{equation}
\zeta_s=\sum_{k=1}^K\sum_{i=1}^n\big(\mathbf{E}[z_{ik}|\mathbf{T};\Theta^*]\cdot m_{si}\big).
\end{equation}
Hence, we obtain the updating equation \eqref{update.iprob}:
\begin{equation}
P_{sk}=\frac{\sum_{i=1}^n\big(\mathbf{E}[z_{ik}|\mathbf{T};\Theta^*]\cdot m_{si}\big)}
            {\sum_{k=1}^K\sum_{i=1}^n\big(\mathbf{E}[z_{ik}|\mathbf{T};\Theta^*]\cdot m_{si}\big)}.
\end{equation}
The updating equation \eqref{update.pi} for $(\pi_1,\pi_2,\ldots,\pi_K)$ can be derived in a similar manner; the actual derivation is omitted.

\subsection{$\mathbf{E}[\log\mathcal{L}(\mathbf{T,Z})|\mathbf{T};\Theta^*]$ under model simplifications \eqref{sim1}-\eqref{sim2}}
\label{appdx:condEdetail-simp}

In Section 5, we introduced further simplifications to our Continuous-time SBM, namely \eqref{sim1} and \eqref{sim2}, before applying it to analyze basketball games. Here, we provide details about the changes to some of the components \eqref{estep1}-\eqref{estep5} for $\mathbf{E}[\log\mathcal{L}(\mathbf{T,Z})|\mathbf{T};\Theta^*]$ as a result of these simplifications. The components \eqref{estep1} and \eqref{estep5} do not involve any rate functions, so they remain the same; whereas the components \eqref{estep2}-\eqref{estep4} now become
\begin{align}
&\sum_{1\leq i\neq j\leq n}\mathbf{E}\Big[\log\mathcal{L}^{P_1}(\mathbf{T}_{ij}|\mathbf{Z})\Big|\mathbf{T};\Theta^*\Big] \label{estep22} \\
&=\sum_{1\leq i\neq j\leq n}\sum_{k=1}^K\sum_{l=1}^K\Big(\mathbf{E}[z_{ik}z_{jl}|\mathbf{T};\Theta^*]\cdot \big(\sum_{h=1}^{m_{ij}}\log\lambda_{k}(t_{ijh})+m_{ij}\log P_{kl}\big)\Big)  \nonumber \\
& \qquad \qquad \qquad \qquad  \quad - \sum_{1\leq i\neq j\leq n}\sum_{k=1}^K\sum_{l=1}^K\mathbf{E}\Big[z_{ik}z_{jl}\cdot\sum_{h=1}^{m_{ij}}\log{G_l^{ijh}}\Big|\mathbf{T};\Theta^*\Big], \nonumber
\end{align}
\begin{align}
&\sum_{i=1}^n \mathbf{E}\Big[\log\mathcal{L}^{P_2}(\mathbf{T}_i|\mathbf{Z})\Big|\mathbf{T};\Theta^* \Big] \label{estep23} \\
&=-\sum_{i=1}^n \sum_{k=1}^K\sum_{l=1}^K\sum_{h=1}^{M_i}\Big(\mathbf{E}\Big[z_{ik}I(G_l^{ih}>0)\Big|\mathbf{T};\Theta^* \Big]\cdot P_{kl} \cdot \int_{t_{ih}^-}^{t_{ih}}\lambda_{k}(t) \mathrm{d}t\Big), \nonumber
\end{align}
and
\begin{multline}
 \sum_{i=1}^{n}\sum_{a\in\mathcal{A}} \mathbf{E}\Big[\log\mathcal{L}^O(\mathbf{T}_{ia}|\mathbf{Z})\Big|\mathbf{T};\Theta^* \Big]
=\sum_{i=1}^{n}\sum_{a\in\mathcal{A}}\sum_{k=1}^K\Big(\mathbf{E}[z_{ik}|\mathbf{T};\Theta^* ]\cdot \\
 \big(\sum_{h=1}^{m_{ia}}\log\lambda_{k}(t_{iah})+m_{ia}\log P_{ka}-P_{ka}\cdot\sum_{h=1}^{M_i}\int_{t_{ih}^-}^{t_{ih}}\lambda_{k}(t)\mathrm{d}t\big)\Big).
 \label{estep24}
\end{multline}

\subsection{Analytic updates of $P_{kl}, P_{ka}$ under model simplifications \eqref{sim1}-\eqref{sim2}} \label{appen4}
\label{appdx:prob-updates-simp}

Recall that, under model simplifications \eqref{sim1}-\eqref{sim2}, the constraint on these transition probabilities is given by \eqref{const}:
\begin{equation}
\sum_{l=1}^{K} P_{kl}+\sum_{a\in\mathcal{A}} P_{ka}=1, \mbox{for any } k=1,2,\ldots,K.
\end{equation}
Again, we introduce Lagrange multiplier $\zeta_k$ for $k=1,2,\ldots,K$. Combining the terms from \eqref{estep22}-\eqref{estep24} that involve these transition probabilities with the constraint above, we obtain the Lagrangian function,
\begin{align}
&\sum_{1\leq i\neq j\leq n}\sum_{k=1}^K\sum_{l=1}^K\Big(\mathbf{E}[z_{ik}z_{jl}|\mathbf{T};\Theta^*]\cdot m_{ij}\cdot \log P_{kl}\Big) \label{trans.p} \\
&-\sum_{i=1}^n \sum_{k=1}^K\sum_{l=1}^K\sum_{h=1}^{M_i}\Big(\mathbf{E}\Big[z_{ik}I(G_l^{ih}>0)\Big|\mathbf{T};\Theta^* \Big]\cdot P_{kl} \cdot \int_{t_{ih}^-}^{t_{ih}}\lambda_{k}(t) \mathrm{d}t\Big) \nonumber \\
& +\sum_{i=1}^{n}\sum_{a\in\mathcal{A}}\sum_{k=1}^K\Big(\mathbf{E}[z_{ik}|\mathbf{T};\Theta^* ] \cdot \big(m_{ia}\cdot \log P_{ka}-P_{ka}\cdot\sum_{h=1}^{M_i}\int_{t_{ih}^-}^{t_{ih}}\lambda_{k}(t)\mathrm{d}t\big)\Big)\nonumber \\
& \qquad \qquad \qquad \qquad \qquad \qquad -\sum_{k=1}^K\zeta_k\cdot \bigg(\sum_{l=1}^KP_{kl}+\sum_{a\in\mathcal{A}}P_{ka}-1\bigg). \nonumber
\end{align}
Differentiating with respect to each $P_{kl}$, $P_{ka}$ and setting the the derivatives to zero, we get
\begin{align}
&\frac{\sum_{1\leq i\neq j\leq n}\Big(\mathbf{E}[z_{ik}z_{jl}|\mathbf{T};\Theta^*]\cdot m_{ij}\Big)}{P_{kl}} \\
&-\sum_{i=1}^n \sum_{h=1}^{M_i}\Big(\mathbf{E}\Big[z_{ik}I(G_l^{ih}>0)\Big|\mathbf{T};\Theta^* \Big]\cdot \int_{t_{ih}^-}^{t_{ih}}\lambda_{k}(t) \mathrm{d}t\Big)-\zeta_k=0, \nonumber
\end{align}
and
\begin{multline}
\frac{\sum_{i=1}^n\Big(\mathbf{E}[z_{ik}|\mathbf{T};\Theta^*]\cdot m_{ia}\Big)}{P_{ka}} \\
-\sum_{i=1}^n \sum_{h=1}^{M_i}\Big(\mathbf{E}[z_{ik}|\mathbf{T};\Theta^*]\cdot \int_{t_{ih}^-}^{t_{ih}}\lambda_{k}(t) \mathrm{d}t\Big)
-\zeta_k=0,
\end{multline}
from which we can solve for the transition probabilities:
\begin{equation}
P_{kl}=\frac{\sum_{1\leq i\neq j\leq n}\Big(\mathbf{E}[z_{ik}z_{jl}|\mathbf{T};\Theta^*]\cdot m_{ij}\Big)}{\sum_{i=1}^n \sum_{h=1}^{M_i}\Big(\mathbf{E}\Big[z_{ik}I(G_l^{ih}>0)\Big|\mathbf{T};\Theta^* \Big]\cdot \int_{t_{ih}^-}^{t_{ih}}\lambda_{k}(t) \mathrm{d}t\Big)+\zeta_k}
\end{equation}
\begin{equation}
P_{ka}=\frac{\sum_{i=1}^n\Big(\mathbf{E}[z_{ik}|\mathbf{T};\Theta^*]\cdot m_{ia}\Big)}{\sum_{i=1}^n \sum_{h=1}^{M_i}\Big(\mathbf{E}[z_{ik}|\mathbf{T};\Theta^*]\cdot \int_{t_{ih}^-}^{t_{ih}}\lambda_{k}(t) \mathrm{d}t\Big) +\zeta_k}
\end{equation}
for $k,l=1,2,\ldots,K$ and $a\in\mathcal{A}$. Each Lagrange multiplier $\zeta_k$ can be solved numerically as the (univariate) root to the equation $\sum_{l=1}^{K}P_{kl}+\sum_{a\in\mathcal{A}}P_{ka}=1$ for each $k$. We do this with the R function \verb!uniroot!.

%% file: Appendix-CI.tex
\section{Confidence Bands for Estimated Rate Functions}
\label{appdxCI}

\begin{figure}[h]
\centering
\includegraphics[width=95mm]{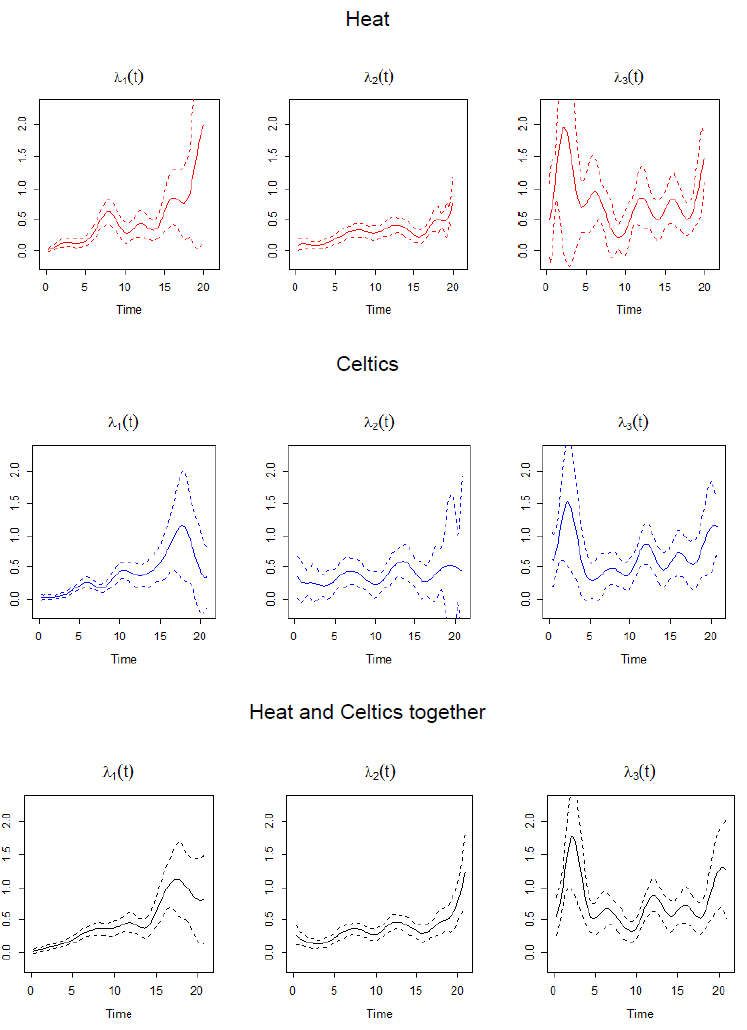}
\caption{\label{Rate.CI}%
Rate functions displayed on top of each other in Figure~\ref{Rate.k3} are displayed here individually with 95\% pointwise confidence bands.}
\end{figure}

It is possible to obtain confidence bands for the estimated rate functions conditional on the cluster labels by calculating the pointwise standard errors using the observed Fisher information matrix and the standard Delta method. As an example, rate functions displayed on top of each other in Figure~\ref{Rate.k3} (to facilitate side-by-side comparison in Section~\ref{sec.HC}) are now displayed individually in Figure~\ref{Rate.CI} with their respective 95\% confidence bands. In all panels of Figure \ref{Rate.CI}, we can see that, as $t\rightarrow24$, the confidence intervals invariably widen. This is because there are fewer transactions as the time approaches the end limit for each play, since many plays end before reaching the full $24$-second limit. Elsewhere, these confidence intervals are narrow enough to suggest that features identified in Figure~\ref{Rate.k3} and discussed in Section~\ref{sec.HC} are unlikely to be merely artifacts due to noise in the data.

%% file: CSBM_REV1-MZ.bbl
\begin{thebibliography}{19}

\bibitem[\protect\citeauthoryear{Airoldi et~al.}{2008}]{Airoldi2008}
\begin{barticle}[author]
\bauthor{\bsnm{Airoldi},~\bfnm{Edoardo~M.}\binits{E.~M.}},
  \bauthor{\bsnm{Blei},~\bfnm{David~M.}\binits{D.~M.}},
  \bauthor{\bsnm{Fienberg},~\bfnm{Stephen~E.}\binits{S.~E.}} \AND
  \bauthor{\bsnm{Xing},~\bfnm{Eric~P.}\binits{E.~P.}}
(\byear{2008}).
\btitle{Mixed Membership Stochastic Blockmodels}.
\bjournal{Journal of Machine Learning Research}
\bvolume{9}
\bpages{1981--2014}.
\end{barticle}
\endbibitem

\bibitem[\protect\citeauthoryear{Bickel and Chen}{2009}]{Bickel2009}
\begin{barticle}[author]
\bauthor{\bsnm{Bickel},~\bfnm{Peter~J.}\binits{P.~J.}} \AND
  \bauthor{\bsnm{Chen},~\bfnm{Aiyou}\binits{A.}}
(\byear{2009}).
\btitle{A nonparametric view of network models and Newman-Girvan and other
  modularities}.
\bjournal{PNAS}
\bvolume{106}
\bpages{21068--21073}.
\end{barticle}
\endbibitem

\bibitem[\protect\citeauthoryear{Cervone et~al.}{2016}]{Bornn2}
\begin{barticle}[author]
\bauthor{\bsnm{Cervone},~\bfnm{Daniel}\binits{D.}},
  \bauthor{\bsnm{D'Amour},~\bfnm{Alex}\binits{A.}},
  \bauthor{\bsnm{Bornn},~\bfnm{Luke}\binits{L.}} \AND
  \bauthor{\bsnm{Goldsberry},~\bfnm{Kirk}\binits{K.}}
(\byear{2016}).
\btitle{A Multireolution Stochastic Process Model for Predicting Basketball
  Possession Outcomes}.
\bjournal{arXiv:1408.0777v3}.
\end{barticle}
\endbibitem

\bibitem[\protect\citeauthoryear{Choi, Wolfe and Airoldi}{2012}]{Choi2012}
\begin{barticle}[author]
\bauthor{\bsnm{Choi},~\bfnm{David~S.}\binits{D.~S.}},
  \bauthor{\bsnm{Wolfe},~\bfnm{Patrick~J.}\binits{P.~J.}} \AND
  \bauthor{\bsnm{Airoldi},~\bfnm{Edoardo~M.}\binits{E.~M.}}
(\byear{2012}).
\btitle{Stochastic Blockmodels with a Growing Number of Classes}.
\bjournal{Biometrika}
\bvolume{99}
\bpages{273-284}.
\end{barticle}
\endbibitem

\bibitem[\protect\citeauthoryear{Cook and Lawless}{2007}]{Cook.book}
\begin{bbook}[author]
\bauthor{\bsnm{Cook},~\bfnm{Richard~J.}\binits{R.~J.}} \AND
  \bauthor{\bsnm{Lawless},~\bfnm{Jerald~F.}\binits{J.~F.}}
(\byear{2007}).
\btitle{The stiatistical analysis of recurrent events}.
\bpublisher{Springer}, \baddress{New York, NY}.
\end{bbook}
\endbibitem

\bibitem[\protect\citeauthoryear{DuBois, Butts and Smyth}{2013}]{UCI}
\begin{barticle}[author]
\bauthor{\bsnm{DuBois},~\bfnm{Christopher}\binits{C.}},
  \bauthor{\bsnm{Butts},~\bfnm{Carter~T.}\binits{C.~T.}} \AND
  \bauthor{\bsnm{Smyth},~\bfnm{Padhraic}\binits{P.}}
(\byear{2013}).
\btitle{Stochastic Blockmodeling of relational event dynamics}.
\bjournal{Proceedings of the 16th International Conference on Artificial
  Intelligence and Statistics (AISTATS)}.
\end{barticle}
\endbibitem

\bibitem[\protect\citeauthoryear{Fewell et~al.}{2012}]{Bbnet}
\begin{barticle}[author]
\bauthor{\bsnm{Fewell},~\bfnm{Jennifer~H.}\binits{J.~H.}},
  \bauthor{\bsnm{Armbruster},~\bfnm{Dieter}\binits{D.}},
  \bauthor{\bsnm{Ingraham},~\bfnm{John}\binits{J.}},
  \bauthor{\bsnm{Petersen},~\bfnm{Alexander}\binits{A.}} \AND
  \bauthor{\bsnm{Waters},~\bfnm{James~S.}\binits{J.~S.}}
(\byear{2012}).
\btitle{Basketball Teams as Strategic Networks}.
\bjournal{PLoS ONE}
\bvolume{7}
\bpages{849--911}.
\end{barticle}
\endbibitem

\bibitem[\protect\citeauthoryear{Ho, Song and Xing}{2011}]{Ho2011}
\begin{barticle}[author]
\bauthor{\bsnm{Ho},~\bfnm{Qirong}\binits{Q.}},
  \bauthor{\bsnm{Song},~\bfnm{Le}\binits{L.}} \AND
  \bauthor{\bsnm{Xing},~\bfnm{Eric~P.}\binits{E.~P.}}
(\byear{2011}).
\btitle{Evolving Cluster Mixed-Membership Blockmodel for Time-Varying
  Networks}.
\bjournal{Proceedings of the 14th International Conference on Artifical
  Intelligence and Statistics (AISTATS)}.
\end{barticle}
\endbibitem

\bibitem[\protect\citeauthoryear{Holland, Laskey and
  Leinhardt}{1983}]{firststep}
\begin{barticle}[author]
\bauthor{\bsnm{Holland},~\bfnm{Paul~W.}\binits{P.~W.}},
  \bauthor{\bsnm{Laskey},~\bfnm{Kathryn~Blackmond}\binits{K.~B.}} \AND
  \bauthor{\bsnm{Leinhardt},~\bfnm{Samuel}\binits{S.}}
(\byear{1983}).
\btitle{Stochastic Blockmodels: First Steps}.
\bjournal{Social Networks}
\bvolume{5}
\bpages{109-137}.
\end{barticle}
\endbibitem

\bibitem[\protect\citeauthoryear{Karrer and Newman}{2011}]{Karrer2011}
\begin{barticle}[author]
\bauthor{\bsnm{Karrer},~\bfnm{Brian}\binits{B.}} \AND
  \bauthor{\bsnm{Newman},~\bfnm{M.~E.~J.}\binits{M.~E.~J.}}
(\byear{2011}).
\btitle{Stochastic blockmodels and community structure in networks}.
\bjournal{Physical Review E}
\bvolume{83}
\bpages{016107}.
\end{barticle}
\endbibitem

\bibitem[\protect\citeauthoryear{Oliver}{2004}]{Oliver}
\begin{bbook}[author]
\bauthor{\bsnm{Oliver},~\bfnm{Dean}\binits{D.}}
(\byear{2004}).
\btitle{Basketball On Paper: Rules and Tools for Performance Analysis}.
\bpublisher{Potomac Books, Inc.}, \baddress{Dulles, Virginia}.
\end{bbook}
\endbibitem

\bibitem[\protect\citeauthoryear{Rohe, Chatterjee and Yu}{2011}]{Rohe2011}
\begin{barticle}[author]
\bauthor{\bsnm{Rohe},~\bfnm{Karl}\binits{K.}},
  \bauthor{\bsnm{Chatterjee},~\bfnm{Sourav}\binits{S.}} \AND
  \bauthor{\bsnm{Yu},~\bfnm{Bin}\binits{B.}}
(\byear{2011}).
\btitle{Spectral Clustering and the High-dimensional Stochastic Blockmodel}.
\bjournal{The Annals of Statistics}
\bvolume{39}
\bpages{1878-1915}.
\end{barticle}
\endbibitem

\bibitem[\protect\citeauthoryear{Shafiei and Chipman}{2010}]{Hugh2010}
\begin{barticle}[author]
\bauthor{\bsnm{Shafiei},~\bfnm{Mahdi}\binits{M.}} \AND
  \bauthor{\bsnm{Chipman},~\bfnm{Hugh}\binits{H.}}
(\byear{2010}).
\btitle{Mixed-Membership Stochastic Block-Models for Transactional Networks}.
\bjournal{Proceedings of the International Conference on Data Mining}.
\end{barticle}
\endbibitem

\bibitem[\protect\citeauthoryear{Shea and Baker}{2013}]{Shea}
\begin{bbook}[author]
\bauthor{\bsnm{Shea},~\bfnm{Stephen~M.}\binits{S.~M.}} \AND
  \bauthor{\bsnm{Baker},~\bfnm{Christopher~E.}\binits{C.~E.}}
(\byear{2013}).
\btitle{Basketball Analytics: Objective and Efficient Strategies for
  Understanding How Teams Win}.
\bpublisher{Advanced Metrics, LLC}, \baddress{Lake St. Louis, MO}.
\end{bbook}
\endbibitem

\bibitem[\protect\citeauthoryear{Snijders and Nowicki}{1997}]{Snijders1997}
\begin{barticle}[author]
\bauthor{\bsnm{Snijders},~\bfnm{Tom A.~B.}\binits{T.~A.~B.}} \AND
  \bauthor{\bsnm{Nowicki},~\bfnm{Krzysztof}\binits{K.}}
(\byear{1997}).
\btitle{Estimation and Prediction for Stochastic Blockmodels for Graphs with
  Latent Block Structure}.
\bjournal{Journal of Classification}
\bvolume{14}
\bpages{75--100}.
\end{barticle}
\endbibitem

\bibitem[\protect\citeauthoryear{Vu et~al.}{2011}]{Hunter}
\begin{barticle}[author]
\bauthor{\bsnm{Vu},~\bfnm{Duy~Q.}\binits{D.~Q.}},
  \bauthor{\bsnm{Asuncion},~\bfnm{Arthur~U.}\binits{A.~U.}},
  \bauthor{\bsnm{Hunter},~\bfnm{David~R.}\binits{D.~R.}} \AND
  \bauthor{\bsnm{Smyth},~\bfnm{Padhraic}\binits{P.}}
(\byear{2011}).
\btitle{Continuous-Time Regression Models for Longitudinal Networks}.
\bjournal{Advances in Neural Information Processing Systems}.
\end{barticle}
\endbibitem

\bibitem[\protect\citeauthoryear{Wang and Wong}{1987}]{wangwong}
\begin{barticle}[author]
\bauthor{\bsnm{Wang},~\bfnm{Yuchung~J.}\binits{Y.~J.}} \AND
  \bauthor{\bsnm{Wong},~\bfnm{George~Y.}\binits{G.~Y.}}
(\byear{1987}).
\btitle{Stochastic Blockmodels for Directed Graphs}.
\bjournal{Journal of the American Statistical Association}
\bvolume{82}
\bpages{8-19}.
\end{barticle}
\endbibitem

\bibitem[\protect\citeauthoryear{Xu and Hero}{2014}]{XuHero2014}
\begin{barticle}[author]
\bauthor{\bsnm{Xu},~\bfnm{K.~S.}\binits{K.~S.}} \AND
  \bauthor{\bsnm{Hero},~\bfnm{A.~O.}\binits{A.~O.}}
(\byear{2014}).
\btitle{Dynamic Stochastic Blockmodels for Time-Evolving Social Networks}.
\bjournal{IEEE Journal of Selected Topics in Signal Processing}
\bvolume{8}
\bpages{552-562}.
\end{barticle}
\endbibitem

\bibitem[\protect\citeauthoryear{Zhao, Levina and Zhu}{2012}]{Zhao2012}
\begin{barticle}[author]
\bauthor{\bsnm{Zhao},~\bfnm{Yunpeng}\binits{Y.}},
  \bauthor{\bsnm{Levina},~\bfnm{Elizaveta}\binits{E.}} \AND
  \bauthor{\bsnm{Zhu},~\bfnm{Ji}\binits{J.}}
(\byear{2012}).
\btitle{Consistency of Community Detection in Networks under Degree-Corrected
  Stoachastic Block Models}.
\bjournal{The Annals of Statistics}
\bvolume{40}
\bpages{2266-2292}.
\end{barticle}
\endbibitem

\end{thebibliography}
